\documentclass[iop,apj,twocolappendix]{emulateapj}

\usepackage{natbib}
\usepackage{amsmath,amssymb}
\usepackage{apjfonts}

\usepackage[usenames]{xcolor}
\definecolor{mplblue}{HTML}{1F77B4}
\definecolor{mplorange}{HTML}{FF7F0E}
\definecolor{mplgreen}{HTML}{2CA02C}
\definecolor{mplred}{HTML}{D62728}

\usepackage[breaklinks,plainpages=false,colorlinks=true,anchorcolor=cyan,linkcolor=mplred,citecolor=cyan,urlcolor=magenta]{hyperref}

\shorttitle{Ephemeris uncertainties vs.\ pulsar timing arrays}
\shortauthors{Vallisneri et al.}

\begin{document}


\title{Modeling the uncertainties of solar-system ephemerides for robust gravitational-wave searches with pulsar timing arrays}


\author{
M.~Vallisneri\altaffilmark{1,2},
S.~R.~Taylor\altaffilmark{3,1,2},
J.~Simon\altaffilmark{1,2},
W.~M.~Folkner\altaffilmark{1},
R.~S.~Park\altaffilmark{1},
C.~Cutler\altaffilmark{1,2},
J.~A.~Ellis\altaffilmark{4},
T.~J.~W.~Lazio\altaffilmark{1,2},
S.~J.~Vigeland\altaffilmark{5},
K.~Aggarwal\altaffilmark{6,7},
Z.~Arzoumanian\altaffilmark{8},
P.~T.~Baker\altaffilmark{9,7},
A.~Brazier\altaffilmark{10,11},
P.~R.~Brook\altaffilmark{6,7},
S.~Burke-Spolaor\altaffilmark{6,7},
S.~Chatterjee\altaffilmark{10},
J.~M.~Cordes\altaffilmark{10},
N.~J.~Cornish\altaffilmark{12},
F.~Crawford\altaffilmark{13},
H.~T.~Cromartie\altaffilmark{14},
K.~Crowter\altaffilmark{15},
M.~DeCesar\altaffilmark{$\dagger$,16},
P.~B.~Demorest\altaffilmark{17},
T.~Dolch\altaffilmark{18},
R.~D.~Ferdman\altaffilmark{19},
E.~C.~Ferrara\altaffilmark{20},
E.~Fonseca\altaffilmark{21},
N.~Garver-Daniels\altaffilmark{6,7},
P.~Gentile\altaffilmark{6,7},
D.~Good\altaffilmark{15},
J.~S.~Hazboun\altaffilmark{$\dagger$,22},
A.~M.~Holgado\altaffilmark{23},
E.~A.~Huerta\altaffilmark{23},
K.~Islo\altaffilmark{5},
R.~Jennings\altaffilmark{10},
G.~Jones\altaffilmark{24},
M.~L.~Jones\altaffilmark{5},
D.~L.~Kaplan\altaffilmark{5},
L.~Z.~Kelley\altaffilmark{25},
J.~S.~Key\altaffilmark{22},
M.~T.~Lam\altaffilmark{26,27,6,7},
L.~Levin\altaffilmark{28},
D.~R.~Lorimer\altaffilmark{6,7},
J.~Luo\altaffilmark{21},
R.~S.~Lynch\altaffilmark{29},
D.~R.~Madison\altaffilmark{$\dagger$,6,7},
M.~A.~McLaughlin\altaffilmark{6,7},
S.~T.~McWilliams\altaffilmark{6,7},
C.~M.~F.~Mingarelli\altaffilmark{30,31},
C.~Ng\altaffilmark{32},
D.~J.~Nice\altaffilmark{16},
T.~T.~Pennucci\altaffilmark{$\dagger$,33},
N.~S.~Pol\altaffilmark{6,7},
S.~M.~Ransom\altaffilmark{14,34},
P.~S.~Ray\altaffilmark{35},
X.~Siemens\altaffilmark{36,5},
R.~Spiewak\altaffilmark{5,37},
I.~H.~Stairs\altaffilmark{15},
D.~R.~Stinebring\altaffilmark{38},
K.~Stovall\altaffilmark{$\dagger$,17},
J.~K.~Swiggum\altaffilmark{$\dagger$,5},
R.~van~Haasteren\altaffilmark{1},
C.~A.~Witt\altaffilmark{6,7},
W.~W.~Zhu\altaffilmark{39}\\
(The NANOGrav Collaboration)\altaffilmark{$\star$}}

\affil{$\star$Author order alphabetical by surname}

\affil{$^{1}$Jet Propulsion Laboratory, California Institute of Technology, 4800 Oak Grove Drive, Pasadena, CA 91109, USA}
\affil{$^{2}$Theoretical AstroPhysics Including Relativity (TAPIR), MC 350-17, California Institute of Technology, Pasadena, CA 91125, USA}
\affil{$^{3}$Department of Physics and Astronomy, Vanderbilt University, 2301 Vanderbilt Place, Nashville, TN 37235, USA}
\affil{$^{4}$Infinia ML, 202 Rigsbee Avenue, Durham, NC 27701, USA}
\affil{$^{5}$Center for Gravitation, Cosmology and Astrophysics, Department of Physics, University of Wisconsin-Milwaukee,\\ P.O.~Box 413, Milwaukee, WI 53201, USA}
\affil{$^{6}$Department of Physics and Astronomy, West Virginia University, P.O.~Box 6315, Morgantown, WV 26506, USA}
\affil{$^{7}$Center for Gravitational Waves and Cosmology, West Virginia University, Chestnut Ridge Research Bldg., Morgantown, WV 26505, USA}
\affil{$^{8}$X-Ray Astrophysics Laboratory, NASA Goddard Space Flight Center, Code 662, Greenbelt, MD 20771, USA}
\affil{$^{9}$Department of Physics and Astronomy, Widener University, One University Place, Chester, PA 19013, USA}
\affil{$^{10}$Department of Astronomy, Cornell University, Ithaca, NY 14853, USA}
\affil{$^{11}$Cornell Center for Advanced Computing, Ithaca, NY 14853, USA}
\affil{$^{12}$Department of Physics, Montana State University, Bozeman, MT 59717, USA}
\affil{$^{13}$Department of Physics and Astronomy, Franklin \& Marshall College, P.O.~Box 3003, Lancaster, PA 17604, USA}
\affil{$^{14}$University of Virginia, Department of Astronomy, P.O.~Box 400325, Charlottesville, VA 22904, USA}
\affil{$^{15}$Department of Physics and Astronomy, University of British Columbia, 6224 Agricultural Road, Vancouver, BC V6T 1Z1, Canada}
\affil{$^{16}$Department of Physics, Lafayette College, Easton, PA 18042, USA}
\affil{$^{17}$National Radio Astronomy Observatory, 1003 Lopezville Rd., Socorro, NM 87801, USA}
\affil{$^{18}$Department of Physics, Hillsdale College, 33 E.~College Street, Hillsdale, MI 49242, USA}
\affil{$^{19}$Department of Physics, University of East Anglia, Norwich, UK}
\affil{$^{20}$NASA Goddard Space Flight Center, Greenbelt, MD 20771, USA}
\affil{$^{21}$Department of Physics, McGill University, 3600  University St., Montreal, QC H3A 2T8, Canada}
\affil{$^{22}$University of Washington Bothell, 18115 Campus Way NE, Bothell, WA 98011, USA}
\affil{$^{23}$NCSA and Department of Astronomy, University of Illinois at Urbana-Champaign, Urbana, IL 61801, USA}
\affil{$^{24}$Department of Physics, Columbia University, New York, NY 10027, USA}
\affil{$^{25}$Center for Interdisciplinary Exploration and Research in Astrophysics (CIERA), Northwestern University, Evanston, IL 60208}
\affil{$^{26}$School of Physics and Astronomy, Rochester Institute of Technology, Rochester, NY 14623, USA}
\affil{$^{27}$Laboratory for Multiwavelength Astronomy, Rochester Institute of Technology, Rochester, NY 14623, USA}
\affil{$^{28}$Jodrell Bank Centre for Astrophysics, University of Manchester, Manchester, M13 9PL, United Kingdom}
\affil{$^{29}$Green Bank Observatory, P.O.~Box 2, Green Bank, WV 24944, USA}
\affil{$^{30}$Department of Physics, University of Connecticut, 196 Auditorium Road, U-3046, Storrs, CT 06269-3046, USA}
\affil{$^{31}$Center for Computational Astrophysics, Flatiron Institute, 162 Fifth Avenue, New York, NY 10010, USA}
\affil{$^{32}$Dunlap Institute for Astronomy and Astrophysics, University of Toronto, 50 St. George St., Toronto, ON M5S 3H4, Canada}
\affil{$^{33}$Hungarian Academy of Sciences MTA-ELTE Extragalactic Astrophysics Research Group, Institute of Physics, E\"{o}tv\"{o}s Lor\'{a}nd University, P\'{a}zm\'{a}ny {P.s.} 1/A, 1117 Budapest, Hungary}
\affil{$^{34}$National Radio Astronomy Observatory, 520 Edgemont Road, Charlottesville, VA 22903, USA}
\affil{$^{35}$Naval Research Laboratory, Washington DC 20375, USA}
\affil{$^{36}$Department of Physics, Oregon State University, Corvallis, OR 97331, USA}
\affil{$^{37}$Centre for Astrophysics and Supercomputing, Swinburne University of Technology, PO Box 218, Hawthorn, VIC 3122, Australia}
\affil{$^{38}$Department of Physics and Astronomy, Oberlin College, Oberlin, OH 44074, USA}
\affil{$^{39}$CAS Key Laboratory of FAST, Chinese Academy of Science, Beijing 100101, China}

\affil{$^{\dagger}$NANOGrav Physics Frontiers Center Postdoctoral Fellow}
\email[Corresponding author email: ]{Michele.Vallisneri@jpl.nasa.gov}




\begin{abstract}
The regularity of pulsar emissions becomes apparent once we reference the pulses' times of arrivals to the inertial rest frame of the solar system. It follows that errors in the determination of Earth's position with respect to the solar-system barycenter can appear as a time-correlated bias in pulsar-timing residual time series, affecting the searches for low-frequency gravitational waves performed with pulsar timing arrays.
Indeed, recent array datasets yield different gravitational-wave background upper limits and detection statistics when analyzed with different solar-system ephemerides. Crucially, the ephemerides do not generally provide usable error representations.
In this article we describe the motivation, construction, and application of a physical model of solar-system ephemeris uncertainties, which focuses on the degrees of freedom (Jupiter's orbital elements) most relevant to gravitational-wave searches with pulsar timing arrays.
This model, \textsc{BayesEphem}, was used to derive ephemeris-robust results in NANOGrav's 11-yr stochastic-background search, and it provides a foundation for future searches by NANOGrav and other consortia.
The analysis and simulations reported here suggest that ephemeris modeling reduces the gravitational-wave sensitivity of the 11-yr dataset; and that this degeneracy will vanish with improved ephemerides and with the longer pulsar timing datasets that will become available in the near future.
\end{abstract}


\maketitle


\section{Introduction}
\label{sec:intro}

Pulsar timing exploits the remarkable regularity of millisecond-pulsar emissions to extract accurate system parameters from time-of-arrival (TOA) datasets \citep{2012hpa..book.....L},
by fitting precise \emph{timing models} that account for all pulse delays and advances, from generation near the neutron stars to detection at the radiotelescopes \citep{2013CQGra..30v4001L}.
The largest time-dependent term in the model is the \emph{R{\o}mer delay} \citep{Roemer1676} due to the motion of Earth around the solar-system barycenter (SSB), with magnitude $\sim 500$ s. Solar-system ephemerides (SSEs), such as those produced by the Jet Propulsion Laboratory (JPL; see \citealt{2009IPNPR.178C...1F,2014IPNPR.196C...1F,de435,de436,de438}), are used to convert observatory TOAs to the notional coordinate time of an inertial frame centered at the SSB.
It follows that errors in our estimate of Earth's trajectory around the SSB produce a time-dependent bias in the TOAs.

Throughout many years of pulsar-timing studies and discoveries, SSE errors were always considered negligible compared to all other sources of noise and uncertainty \citep{1990ApJ...361..300F,2006MNRAS.372.1549E}.
TOAs are now being collected with ever greater time spans and timing precisions, especially so by the \emph{pulsar-timing-array} (PTA) collaborations seeking to detect gravitational waves (GWs) as correlated residuals (i.e., TOAs minus timing model) in multi-pulsar datasets \citep{saz78,det79,fb90,ml13,dcl+16,h13,v+16}.
The estimated magnitude of the GW signature is $\sim 100$ ns or less, leading to the recent suggestion that SSE errors could measurably bias GW results obtained from these datasets \citep{2016MNRAS.455.4339T,2019ApJ...876...55R}.
%
\begin{figure*}[t]
\begin{center}
    \includegraphics[width=0.85\textwidth]{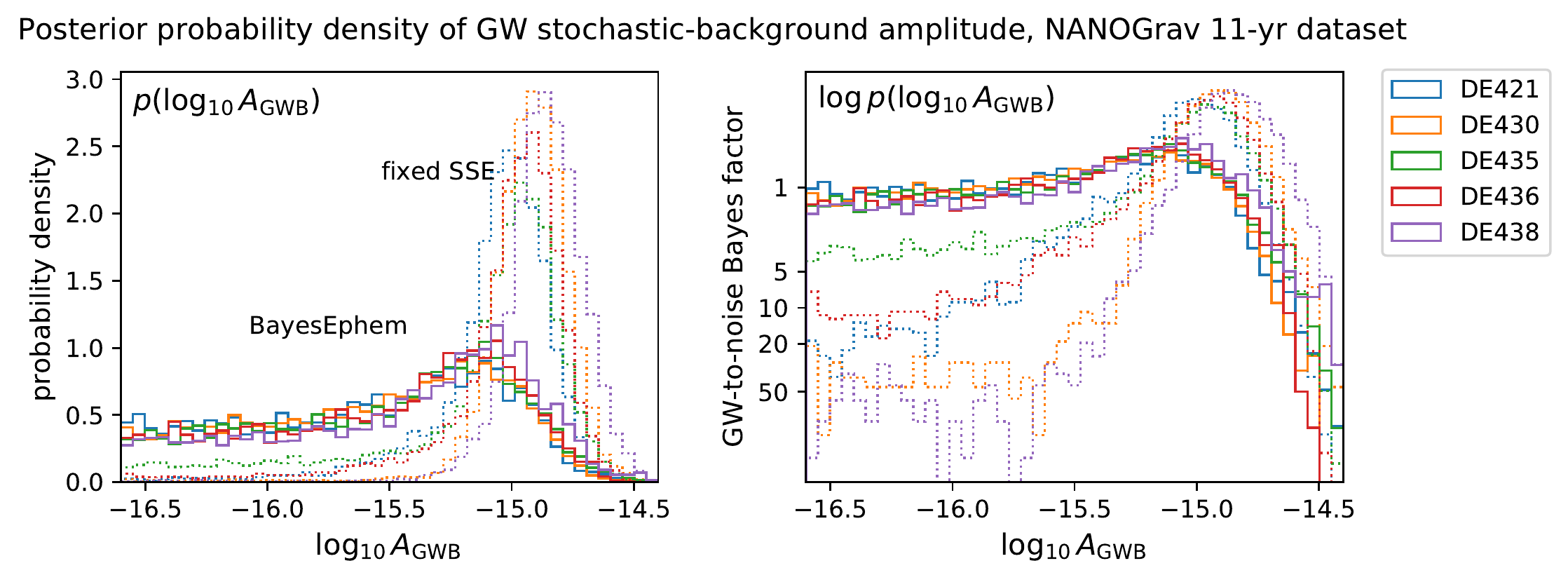}
\end{center}
\vspace{-12pt}
    \caption{(\textbf{left}) Bayesian posteriors for the amplitude $A_\mathrm{GWB}$ (at $f = \mathrm{yr}^{-1}$) of the GW stochastic background, modeled as a fixed power law with slope $\gamma = 11/3$, as appropriate for a population of inspiraling and GW-emitting supermassive black-hole binaries.
    The posteriors are computed for the NANOGrav 11-year dataset using individual JPL ephemerides (dotted lines), and \textsc{BayesEphem} (solid lines). The GW model (``model 2A'' in \citealt{2018ApJ...859...47A}) does not include Hellings--Downs correlations, which however modify results only marginally.
    The incorporation of explicit SSE uncertainties into the analysis via \textsc{BayesEphem} leads to substantially lower evidence of a GW background, and to much greater consistency among the different SSEs.
    (\textbf{right}) Same posteriors as in the left panel, shown with a logarithmic vertical scale that can be mapped to approximate GW vs.\ noise-only Bayes factors, by way of the Savage--Dickey formula $P(A_\mathrm{GWB}=0)/P(A_\mathrm{GW}\neq 0) = p(A_\mathrm{GWB}=0|\mathrm{data})/p_\mathrm{prior}(A_\mathrm{GWB}=0)$ \citep{d71}.
    The application of \textsc{BayesEphem} brings all Bayes factors close to unity, indicating no evidence for GWs once SSE uncertainties are taken into consideration.
    Bayes factors and 95\% $A_\mathrm{GWB}$  upper limits calculated from these curves are listed in Tables \ref{tab:bf} and \ref{tab:upper}.
    \label{fig:posteriors}}
\end{figure*}

Within the North American Nanohertz Observatory for Gravitational Waves (NANOGrav), we encountered this bias firsthand in early 2017, when we observed that switching between the more recent SSEs issued by JPL (specifically DE421, DE430, DE435, and DE436: \citealt{2009IPNPR.178C...1F,2014IPNPR.196C...1F,de435,de436}) led to discrepant results (GW upper limits and detection statistics) in the search for a stochastic signal from the population of supermassive black-hole binaries, formed in the centers of galaxies following major-merger events, as performed on the NANOGrav 11-year dataset \citep{2018ApJS..235...37A,2018ApJ...859...47A}.
See Fig.\ \ref{fig:posteriors} for the Bayesian posterior probabilities of the GW-background (GWB) amplitude, as obtained with a range of SSEs. We verified that other recent PTA results \citep{2015Sci...349.1522S,2016ApJ...821...13A} would be similarly affected.

In this article we report on \textsc{BayesEphem}, the physical model of SSE uncertainties that we developed and integrated with the NANOGrav GW analysis so that we could produce SSE-robust results \citep{2018ApJ...859...47A} by marginalizing our Bayesian statistics over the SSE model parameters. 
Our model complements published SSEs, which do not generally include usable time-domain representations of orbit uncertainties.
We adopted the conservative goal of \emph{bridging} the JPL SSEs so that our analysis would yield the same GW-amplitude posteriors, no matter which SSE was used to offset the TOAs initially.
As discussed in Sec.\ \ref{sec:physical}, the crucial element in our approach turns out to be the modeling of uncertainties in Jupiter's orbit, which create apparent SSB motions with periods comparable to the duration of our dataset, and with amplitudes $\sim$ 50 m. These correspond to $\sim$ 170-ns delays, within the GW sensitivity of our PTA.
By marginalizing GW posteriors over a set of SSE correction parameters that include Jupiter's orbital elements, we achieve our bridging criterion for the SSEs adopted in \cite{2018ApJ...859...47A}, and indeed also for the newer DE438 \citep{de438}; see Fig.\ \ref{fig:posteriors}.
By contrast, while we include SSB corrections due to perturbations in the masses of the outer planets \citep{2010ApJ...720L.201C}, we conclude that the current sensitivity of our dataset is insufficient to constrain these masses better than recent spacecraft tracking and Doppler datasets \citep{jh+2000,2018MNRAS.481.5501C,2006AJ....132.2520J,2014AJ....148...76J,2009AJ....137.4322J}.

In our 11-yr analysis \citep{2018ApJ...859...47A}, we declined to adopt an aggressive stance that would have given more credence to the more recent available SSEs (DE435 and DE436), which are based on longer sets of solar-system observations that fully cover the span of our PTA dataset, and on more sophisticated analysis techniques.
The resulting GW-amplitude posteriors imply less evidence for GWs than those obtained with earlier SSEs (see Table \ref{tab:bf}), and intermediate estimates for 95\% amplitude upper limits (see Table \ref{tab:upper}).
Nevertheless, even if DE435 and DE436 are described as having only minor differences \citep{de435,de436}, the resulting posteriors are still at variance---especially so in the low GW-amplitude limit, which affects the Savage--Dickey Bayes ratio used as our GW detection statistic.
Thus, uncertainty modeling remains important even if we concentrate on newer SSEs.

This article is organized as follows. In Sec.\ \ref{sec:sses} we summarize the SSE production, and we describe the history and stated accuracy of the JPL SSEs adopted in our work; in Sec.\ \ref{sec:systematics} we discuss the TOA delays induced by SSE errors, which are partially absorbed by the timing model, and we identify Jupiter and Saturn's orbits as the drivers behind GW-posterior discrepancies; in Sec.\ \ref{sec:physical} we formulate \textsc{BayesEphem}, and give details about its implementation in our PTA data-analysis software, \textsc{Enterprise} \citep{2019ascl.soft12015E}; in Sec.\ \ref{sec:results} we report on the Bayesian posteriors (for the GW amplitude and for orbital-correction parameters) obtained with \textsc{BayesEphem} for NANOGrav's 11-yr dataset, reproducing and expanding the results of \cite{2018ApJ...859...47A};
in Sec.\ \ref{sec:simulations} we present simulations that probe the reduction in GW sensitivity due to \textsc{BayesEphem};
in Sec.\ \ref{sec:othermodels} we discuss other approaches toward SSE uncertainty modeling;
last, in Sec.\ \ref{sec:conclusion} we offer our brief conclusions.  

All computational results presented in this article were obtained with \textsc{Enterprise}  \citep{2019ascl.soft12015E}, used in conjunction with the stochastic sampler \textsc{PTMCMCSampler}\footnote{\href{https://github.com/jellis18/PTMCMCSampler}{https://github.com/jellis18/PTMCMCSampler}} and the pulsar-timing package \textsc{Tempo2} \citep{2012ascl.soft10015H}.
Code that supports all the calculations discussed here and that produces all the figures of this paper is available as a Jupyter notebook at \url{https://github.com/nanograv/11yr_stochastic_analysis}.

\section{Solar-system ephemerides}
\label{sec:sses}

To derive the JPL SSEs\footnote{\href{https://ssd.jpl.nasa.gov/?ephemerides}{https://ssd.jpl.nasa.gov/?ephemerides}} (and similarly for the French INPOP\footnote{\href{https://www.imcce.fr/inpop}{https://www.imcce.fr/inpop}} and Russian EPM\footnote{\href{http://iaaras.ru/en/dept/ephemeris/epm}{http://iaaras.ru/en/dept/ephemeris/epm}}),
the orbits of the Sun, the planets, and a large number of asteroids are fit to heterogeneous datasets collected over the last few decades. Measurement techniques include spacecraft ranging and Doppler tracking, direct planetary radar ranging, very long baseline interferometry of spacecraft, and (for the Moon) the laser ranging of retroreflectors left by the Apollo missions (see \citealt{verma2013} for a review). The masses of minor bodies are also included as fit parameters, while the masses of the planets are held fixed to values determined separately from spacecraft data for each planet \citep{jh+2000,2006AJ....132.2520J,2014AJ....148...76J,2009AJ....137.4322J}.

The parameters of the fit are the initial conditions (``epoch'' positions and velocities) for all the bodies (as well as minor-body masses); from these, orbits are integrated numerically, providing a reference solution used to compute measurement residuals.
The integration is repeated with minor displacements in all fit parameters, yielding variational \emph{partials} for the orbits. The fit parameters are then corrected by finding the linear combination of the partials that minimizes the residuals in least-squares fashion, and the scheme is repeated until the solution converges (see, e.g., \citealt{1983A&A...125..150N}).

The most complex aspect of the process is the modeling of observations for datasets that are both varied in technique and unique to each planet and each new spacecraft \citep{moyer2003}---a Sisyphean task. Because it is difficult to assign realistic uncertainties to many of the measurements (and, perhaps more importantly, to estimate their systematic errors), the \emph{formal} errors and covariances estimated with the least-squares procedure are considered unreliable, and are not published with the best-fit orbits. Instead, orbit accuracy is assessed by analyzing model residuals and by comparing estimates that use different subsets of the data \citep{de434cov,de438}.

In this paper we work with the four JPL SSEs used to analyze the NANOGrav 11-year dataset \citep{2018ApJ...859...47A} as well as the more recent DE438. While we have also investigated some of the SSEs provided by INPOP \citep{inpop13,inpop17}, we find that they are not qualitatively different in terms of their GW constraints.
\begin{description}
\item[DE421 \citep{2009IPNPR.178C...1F}] was published in 2009, based on data through 2007. The orbits of the inner planets are known to sub-km accuracy, those of Jupiter and Saturn to tens of km; Uranus and Neptune not well determined. The axes of the ephemeris are oriented with the International Celestial Reference Frame \citep{Ma_1998} with accuracy $\lesssim 1$ mas.
\item[DE430 \citep{2014IPNPR.196C...1F}] was published in 2014, based on data through 2013. Orbit integration relies on a more sophisticated dynamical model. The Saturn orbit is more accurate thanks to the improved treatment of range measurements to the Cassini spacecraft \citep{PhysRevD.89.102002}.
The axes of the ephemeris are oriented with the 2009 update of the International Celestial Reference Frame, ICRF2 \citep{2015AJ....150...58F} with accuracy $\lesssim 0.2$ mas, which represents the limiting error source for the inner planets, corresponding to orbit uncertainties of a few hundred meters. The Jupiter and Saturn orbits are determined to tens of km; those of Uranus and Neptune (which are constrained mainly by astrometric measurements) to thousands of km. 
\item[DE435 \citep{de435}] was published in 2016; it improves the Saturn orbit using Cassini data through 2015, correcting it by $\sim 1.5$ km. The Jupiter orbit had been updated (by $\sim 50$ km) in the 2015 DE434 ephemeris \citep{de434} by reprocessing data from six spacecraft flybys, and adding data from the New Horizons flyby. In DE435, the Jupiter orbit is tweaked further (by $\sim 20$ km) by reweighting the datasets. These changes are deemed ``consistent'' with the estimated orbit uncertainties \citep{de434,de435}.
\item[DE436 \citep{de436}] was published at the end of 2016; it updates the Jupiter ephemeris (by $\sim 20$ km) for use by the Juno navigation team. The Saturn orbit changes very slightly.
\item[DE438 \citep{de438}] was published in June 2018; it updates the Jupiter ephemeris (by $\sim 10$ km) with Juno measurements (six ranges and four VLBI observations near perijove), and the Saturn ephemeris (by $\sim 1$ km) with reprocessed ranges through the end of the Cassini mission. The accuracy of the Jupiter orbits is deemed ``at least a factor of four better than previous ephemerides,'' viz.\ $\lesssim$ 10 km.
\end{description}

\begin{figure}[t]
    \centering
    \includegraphics[width=\columnwidth]{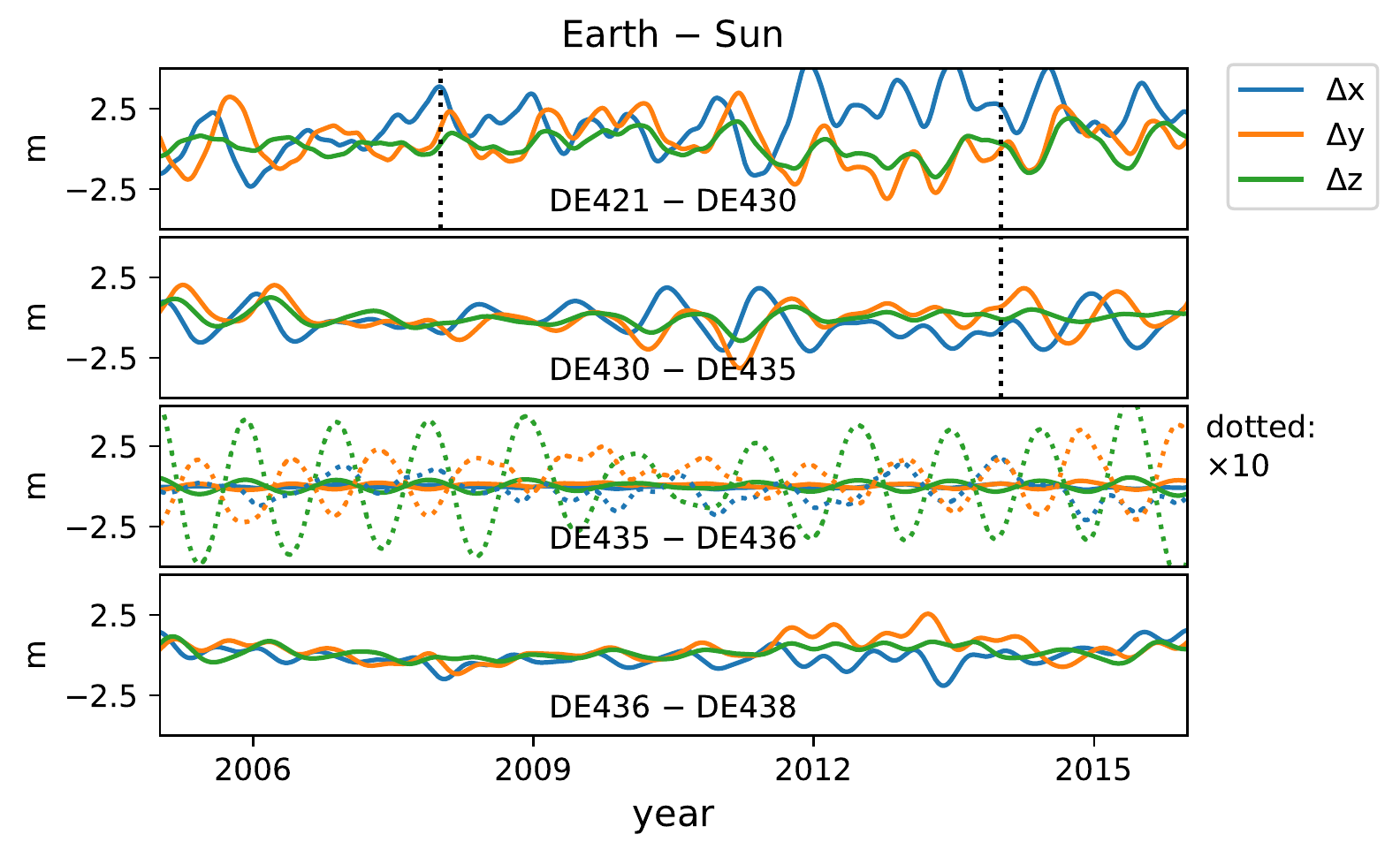}
    \includegraphics[width=\columnwidth]{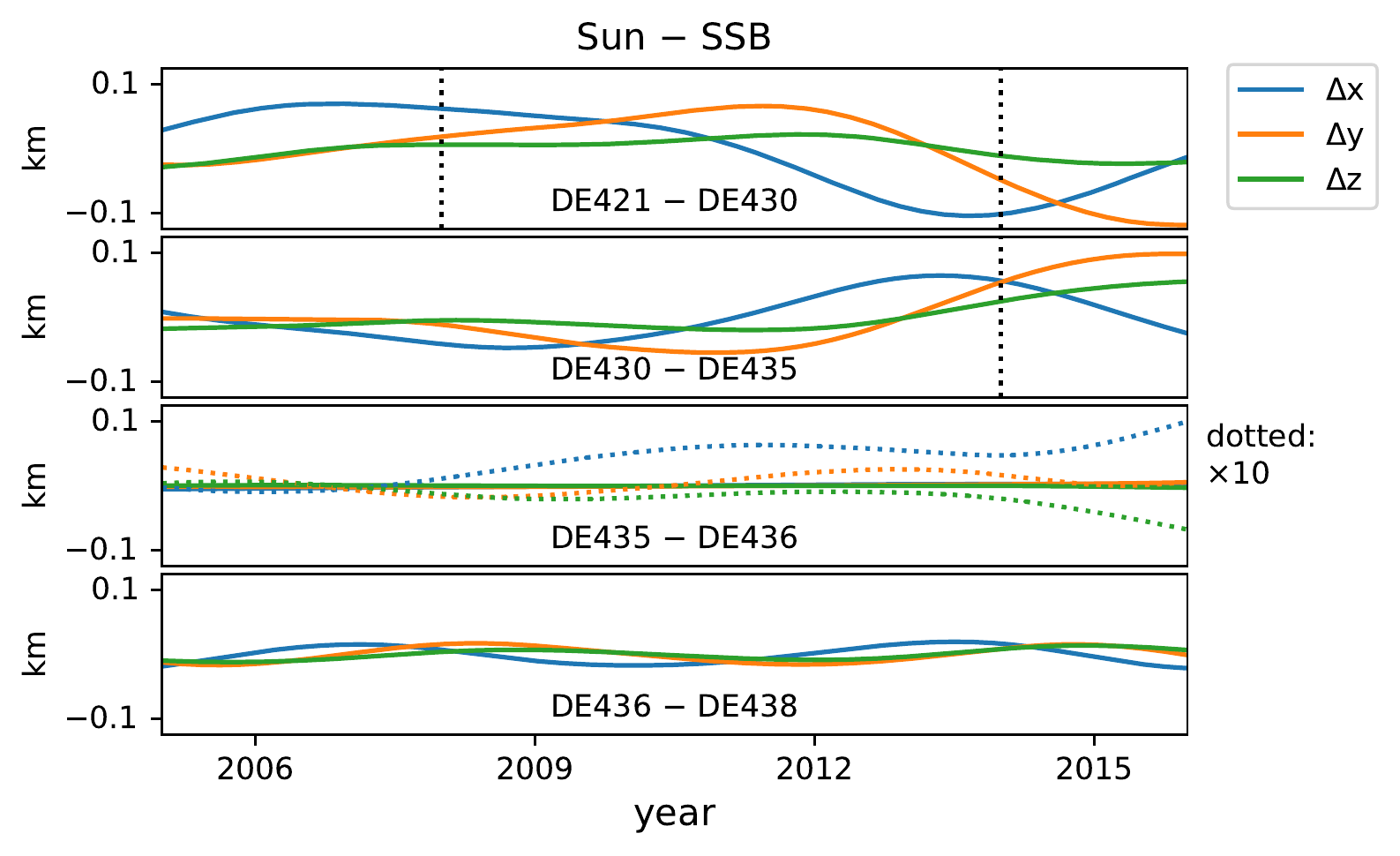}
    \caption{Differences in Earth $-$ Sun position, $\mathbf{r}^{(3)} - \mathbf{r}^\mathrm{Sun}$, and in Sun $-$ SSB position, $\mathbf{r}^\mathrm{Sun} - \mathbf{r}^\mathrm{SSB}$, between pairs of JPL SSEs. Note the difference in the vertical scale of the two panels.
    Earth $-$ SSB plots, not included here, would appear essentially the same as Sun $-$ SSB plots, since the Earth $-$ Sun differences are very small. For reference, 2.5~m $\approx$ 8~ns, and 0.1~km $\approx$ 300~ns, bracketing the range of amplitudes expected for GW signals in PTA datasets. \\
    We plot SSE equatorial-coordinate differences over the approximate span of NANOGrav's 11-yr dataset, and we apply the best-fit frame rotation that minimizes the 3D norm of the difference. Dotted curves are zoomed $\times 10$ vertically. Dotted vertical lines mark the end of the DE421 and DE430 fitted datasets.
    For clarity, we also remove a $\sim 100$-m mean in each coordinate in the Sun--SSB differences; such a constant offset does not affect PTA likelihoods.
    \label{fig:earthsun}}
\end{figure}

Across these ephemerides, the orbit of Earth relative to the Sun is consistent at the $3$-m ($\sim 10$-ns) level, after applying an overall rotation with respect to the International Celestial Reference Frame (within the uncertainties of that ``tie''), and a rotation \emph{rate} about the ecliptic pole\footnote{\label{footnote:rate}This accounts for differences in the estimated semi-major axis of the Earth-Moon--barycenter orbit, which gives rise to a linear rate in estimated ecliptic longitude.}.
However, the orbit of the Sun and therefore the orbit of Earth (both relative to the SSB) match only at the $100$-m ($\sim 300$-ns) level across ephemerides (see Fig.\ \ref{fig:earthsun}). This discrepancy arises from the estimated positions of Jupiter, Saturn, Uranus, and Neptune. Using a simple dynamical model of the SS (i.e., integrating the equations of Newtonian gravity for the eight planets and the Sun), it is easy to show that if we perturb the masses or the orbits of the outer planets, we affect the Sun-to-SSB and Earth-to-SSB trajectories through the resulting redefinition of the SSB, rather than through the very minor changes in the gravitational pull of the outer planets (see also \citealt{2019MNRAS.489.5573G}).

The last few JPL ephemerides focus on Jupiter and Saturn, as motivated by the navigation needs of JPL missions to those planets. In DE421 to DE436, Saturn's orbit is known better than Jupiter's, because the Cassini tracking data is more complete and accurate than was possible for previous spacecraft. In particular, the high-gain antenna of Jupiter orbiter Galileo failed to deploy, leading to low-accuracy measurements. Data from the Juno spacecraft, which has been orbiting Jupiter since July 2016, appears to improve the Jupiter ephemeris substantially \citep{2009IPNPR.178C...1F}.

\section{SSE errors as systematics for pulsar timing arrays}
\label{sec:systematics}

The search for stochastic GW signals with PTAs exploits the distinctive quadrupolar signature in the inter-pulsar correlations of timing-model residuals \citep{hd83}.
Residuals are obtained after applying a chain of corrections that convert the TOA measured at the radiotelescope to the notional emission time in the pulsar system (see, e.g., \citealt{ehm06}):
\begin{equation}
    t_\mathrm{e}^\mathrm{psr} = t_\mathrm{a}^\mathrm{obs} - \Delta_\odot - \Delta_\mathrm{IS} - \Delta_\mathrm{B},
\end{equation}
where $t_\mathrm{e}^\mathrm{psr}$ and $t_\mathrm{a}^\mathrm{obs}$ are the emission and arrival times, $\Delta_\odot$ captures corrections between the observatory and SSB frames, $\Delta_\mathrm{IS}$ describes corrections between the SSB frame and the pulsar-system barycenter, and $\Delta_\mathrm{B}$ models corrections from the pulsar system barycenter to the pulsar frame (which are relevant for binary systems).
Among these terms, $\Delta_\mathrm{IS}$ is very large, but changes very slowly over the duration of pulsar datasets, and thus maps to a constant phase offset. Next comes the R{\o}mer delay \citep{Roemer1676}, corresponding to the light travel time ($\sim 500$ s) between the observatory at $\mathbf{r}^\mathrm{obs}$ and the SSB at $\mathbf{r}^\mathrm{SSB}$:
%
\begin{equation}
\label{eq:Roemer}
\Delta_{\mathrm{R}\odot} =
t_\mathrm{a}^\mathrm{obs} - t_\mathrm{a}^\mathrm{SSB} =
-\frac{\bigl(\mathbf{r}^\mathrm{obs}(t_\mathrm{a}^\mathrm{obs}) - \mathbf{r}^\mathrm{SSB}(t_\mathrm{a}^\mathrm{obs})\bigr) \cdot \hat{\mathbf{p}}}{c},
\end{equation}
where $\hat{\mathbf{p}}$ is the unit vector in the direction of the pulsar. In fact, $\hat{\mathbf{p}}$ is determined from pulsar-timing data mainly through the time dependence of Eq.\ \eqref{eq:Roemer}. 

The errors $\delta \mathbf{r}^\mathrm{obs}$ and $\delta \mathbf{r}^\mathrm{SSB}$ induce systematic TOA delays according to Eq.\ \eqref{eq:Roemer}.
Furthermore, the position of the observatory with respect to Earth's barycenter is known to few-cm (sub-ns) accuracy \citep{ehm06}, so we may write the systematic R{\o}mer-delay error as
\begin{equation}
\label{eq:deltar}
\delta t_{\mathrm{R}\odot} = -\frac{\delta \mathbf{x}^{(3)}(t_\mathrm{a}^\mathrm{obs})\cdot\hat{\mathbf{p}}}{c}.
\end{equation}
where $\mathbf{x}^{(3)} \equiv \mathbf{r}^{(3)} - \mathbf{r}^\mathrm{SSB}$
is the terrestrial barycenter's position in the SSE frame where the SSB is identified with the origin, and $\delta \mathbf{x}^{(3)}(t)$ is the systematic error (a function of time) in the SSE's estimate of $\mathbf{x}^{(3)}(t)$. Here and below we index the solar-system planets from Mercury to Neptune as $(1)$ through $(8)$, so Earth is $(3)$. We will also continue to use $\mathbf{x}$ to refer to vectors in the SSE frame, with origin at the SSB.

The time dependence of the SSE errors is essential to the effect of the ensuing residuals on GW searches in PTA data. Constant offsets, as well as linear and quadratic trends, are absorbed by redefinitions of the timing-model parameters that describe the intrinsic spin evolution of the pulsar; in other words, the TOA likelihood is affected only by $\delta t_{\mathrm{R}\odot}$ minus the best-fitting quadratic polynomial\footnote{In a Bayesian context there is no single best-fitting polynomial when noise parameters are included in the inference, leading to varying weights for the fit. Nevertheless the subtracted features remain largely irrelevant to parameter posteriors.}. Other free parameters in the timing model yield similar subtractions: most important, an angular misalignment between the SSE frame and the ``true'' sidereal frame can be absorbed by a slight coherent displacement in the position of all pulsars; the corresponding TOA corrections are sinusoidal with periods of one sidereal year.

Pulsar timing arrays are most sensitive to GWs with periods of a few to several years, on the order of the total timespan of observations;\footnote{This can be understood by noticing that the dominant source of residual noise (``radiometer'' measurement noise) is white, but TOAs are sensitive to time integrals of GW strain, so the effective GW noise grows with frequency.}
therefore SSE errors in the orbits of the giant planets, which have comparable periods, could be mistaken for GWs.
Correcting the orbit of a planet by the time-dependent vector $\delta \mathbf{r}(t)$ corresponds to offsetting the SSB position by $(m^\mathrm{planet}/m^\mathrm{SS}) \times \delta \mathbf{r}(t)$, where $m^\mathrm{SS}$ is the total mass in the solar system. We then estimate
\begin{equation}
    \begin{aligned}
    \delta x^\mathrm{SSB}(\delta x^{(5)}) & \approx 10^{-3} \!\times\! (50 \, \mathrm{km}) \approx 50 \, \mathrm{m} \approx 170 \, \mathrm{ns}, \\
    \delta x^\mathrm{SSB}(\delta x^{(6)}) & \approx (3 \!\times\! 10^{-4}) \!\times\! (50 \, \mathrm{km}) \approx 15 \, \mathrm{m} \approx 50 \, \mathrm{ns}, \\
    \delta x^\mathrm{SSB}(\delta x^{(7,8)}) & \approx (5 \!\times\! 10^{-5}) \!\times\! (5000 \, \mathrm{km}) \approx 250 \, \mathrm{m} \approx 800 \, \mathrm{ns},
    \end{aligned}
\end{equation}
where the quantities in nanoseconds are light-travel times equivalent to the distances.
While the uncertainties due to Uranus and Neptune are larger, their orbital periods ($P^{(7)} = 84$ yr and $P^{(8)} = 165$ yr) ensure that the corresponding $\delta x^\mathrm{SSB}$ appear as linear or mildly quadratic TOA trends, which are absorbed by timing models (and will continue to be, until PTA datasets approach a century in duration). By contrast, Jupiter and Saturn corrections enter the residuals with timescales ($P^{(5)} = 12$ yr and $P^{(6)} = 29$ yr) comparable to the span of our dataset---just where PTAs are most sensitive to GWs.

Likewise, the absolute location of the SSB is degenerate with the initial phase of the pulses for each pulsar, so we need not worry that the former depends strongly on the set of bodies that are included in each SSE fit. For instance, including trans-Neptunian objects relocates the SSB by $\sim$ 100 km \citep{2014IPNPR.196C...1F}.

\citet{2010ApJ...720L.201C} discuss pulsar-timing's potential to constrain the masses of outer planets. The uncertainties in current IAU best estimates (derived from spacecraft tracking and Doppler studies: \citealt{iaumasses,jh+2000,2006AJ....132.2520J,2014AJ....148...76J,2009AJ....137.4322J}) give rise to R{\o}mer corrections comparable to the orbit errors. To wit, these scale as $(\delta m^\mathrm{planet}/m^\mathrm{SS}) \times \mathbf{r}$, so
\begin{equation}
    \begin{aligned}
    \delta x^\mathrm{SSB}(\delta m^{(5)}) & \approx (1.6 \!\times\! 10^{-11})\!\times\!
    (5.2 \, \mathrm{AU})
    \approx 12 \, \mathrm{m} \approx 40 \, \mathrm{ns}, \\
    \delta x^\mathrm{SSB}(\delta m^{(6)}) & \approx (8.2 \!\times\! 10^{-12}) \!\times\!
    (9.4 \, \mathrm{AU})
    \approx 12 \, \mathrm{m} \approx 40 \, \mathrm{ns}, \\
    \delta x^\mathrm{SSB}(\delta m^{(7)}) & \approx (3.2 \!\times\! 10^{-11}) \!\times\!
    (19 \, \mathrm{AU})
    \approx 90 \, \mathrm{m} \approx 300 \, \mathrm{ns}, \\
    \delta x^\mathrm{SSB}(\delta m^{(8)}) & \approx (8.0 \!\times\! 10^{-11}) \!\times\!
    (30 \, \mathrm{AU})
    \approx 360 \, \mathrm{m} \approx 1.2 \, \mu\mathrm{s}.
    \end{aligned}
\end{equation}

These corrections yield delays with the same periods as the corresponding planets, so again they may be observable in PTA datasets for Jupiter and Saturn, but not for Uranus and Neptune.
%
\begin{figure*}[t]
    \centering
    \includegraphics[width=2\columnwidth]{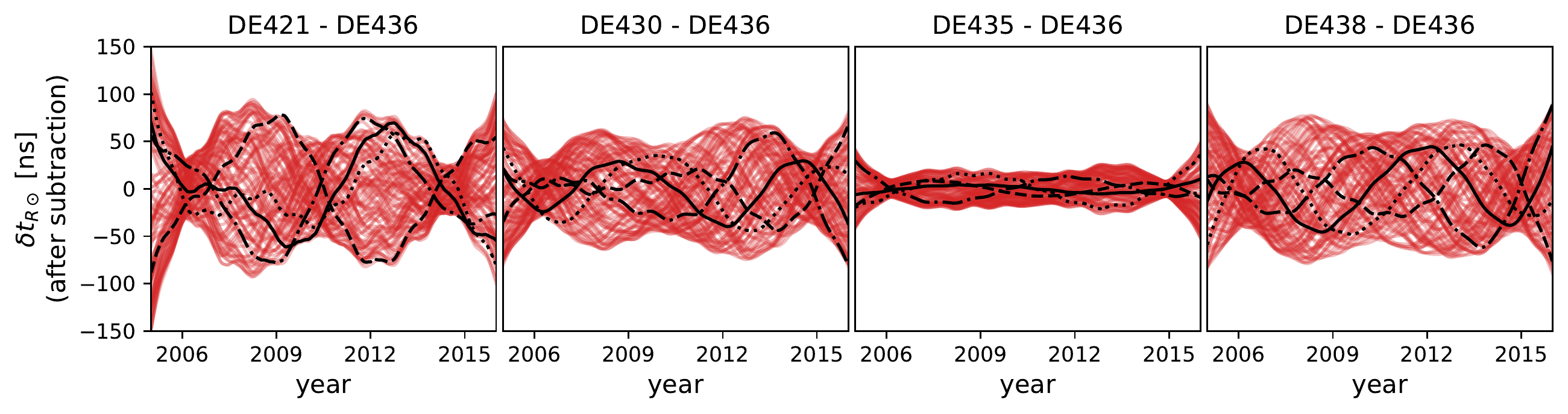}
    \caption{Differences in R{\o}mer delays computed with DE421/DE430/DE435/DE438 and taking DE436 as reference for 100 simulated pulsars randomly distributed across the sky, plotted after subtracting the best-fitting quadratic and yearly sinusoid. The darker curves with various dashing styles allow a comparison of the same pulsars across the three differences. The plot covers the span of the NANOGrav 11-yr dataset.
    For any given pulsar, these differences are reduced to less than 10 ns after subtracting the best-fitting linear combination of \textsc{BayesEphem} partials, dotted into the pulsar position to obtain R{\o}mer delays.
    }
    \label{fig:Roemer}
\end{figure*}

In Fig.\ \ref{fig:Roemer} we show the difference between R{\o}mer delays for 100 simulated pulsars randomly distributed across the sky, as computed with DE421/DE430/DE435/DE438 and with DE436, which is taken as a reference. Thus we are plotting the systematic error that we would introduce using the other SSEs \emph{if} DE436 gave exact orbits. We represent timing-model fits by subtracting the best-fitting quadratics and period--1-yr sinusoids. The plot covers the span of the NANOGrav 11-yr dataset.
The differences reach $\sim 100$ ns with typical periods $\sim 10$--$12$ yr (consistent with a Jovian attribution)---within the sensitivity range and band of GW searches. Indeed, as we see in Fig.\ \ref{fig:posteriors}, the GW-amplitude posteriors for the NANOGrav 11-yr stochastic-background characterization are affected significantly by the choice of SSE.

\section{\textsc{BayesEphem}: a physical model of solar-system-ephemeris uncertainties for gravitational-wave searches in pulsar-timing-array data}
\label{sec:physical}

We set out to address the sensitivity of the NANOGrav stochastic-background analysis to SSE systematics by developing a parametrized physical model of SSE uncertainties (\textsc{BayesEphem}), so that we can compute robust GW-parameter posteriors by marginalizing over the SSE parameters. 
Motivated by the discussion of Secs.\ \ref{sec:sses} and \ref{sec:systematics}, we include the following components:
\begin{itemize} 
\item \textbf{TOA delays generated by corrections to the masses} of Jupiter, Saturn, Uranus, and Neptune, modeled as \cite{2010ApJ...720L.201C}
\begin{equation}
\label{eq:massperturb}
    \delta t_{\mathrm{R}\odot}(\delta m^{(p)}) = -\frac{\delta m^{(p)}}{m^\mathrm{SS}} \frac{\mathbf{x}^{(p)} \cdot \mathbf{p}}{c}.
\end{equation}
We impose normal Bayesian priors on the $\delta m^{(p)}$, with standard deviation equal to the IAU-adopted mass-estimate uncertainties \citep{iaumasses}. For each pulsar dataset, we compute the $\mathbf{x}^{(p)}$ by evaluating the DE436 SSE at the measured TOAs $\{t_i\}$.
%
\item \textbf{TOA delays generated by a rotation rate about the ecliptic pole} (as needed to absorb Sun-to-Earth orbit differences among SSEs),
\begin{equation}
\label{eq:rate}
    \delta t_{\mathrm{R}\odot}(\omega^{\hat{z}}) =
    -\frac{(\mathsf{R}^{\hat{z}}(\delta \theta) \cdot \mathbf{x}^{(3)} - \mathbf{x}^{(3)}) \cdot \mathbf{p}}{c} 
\end{equation}
where $\mathsf{R}^{\hat{z}}(\cdot)$ is the appropriately oriented rotation matrix, and $\delta \theta = \omega^{\hat{z}} (t - t_0)$ is the rotation angle, with $\omega^{\hat{z}}$ the rate and $t_0$ set at the beginning of the NANOGrav dataset.
We impose uniform Bayesian priors on $\omega^{\hat{z}}$ that are commensurate with the rotation rates needed to reduce the difference between the JPL SSEs. 
Given that these rates are very small, we linearize Eq.\ \eqref{eq:rate} with respect to $\omega^{\hat{z}}$; as for $\delta t_{\mathrm{R}\odot}(\delta m^{(p)})$, we compute $\mathbf{x}^{(3)}$ by evaluating the DE436 at the measured TOAs $\{t_i\}$ for each pulsar dataset.

Note that we do not expect this term to affect GW posteriors: both static and uniform rotations of the SSE frame are absorbed in the estimated positions and proper motions of the pulsars. We omit the former altogether, and include $\omega^{\hat{z}}$ as a check.
\item \textbf{TOA delays generated by perturbing Jupiter's average orbital elements}, as given by
\begin{equation}
\label{eq:orbitperturb}
    \delta t_{\mathrm{R}\odot}(\delta a^\mu) = -\delta a^\mu
    \frac{m^{(5)}}{m^\mathrm{SS}} \frac{\partial \mathbf{x}^{(5)}}{\partial a^\mu} \cdot \frac{\mathbf{p}}{c},
\end{equation}
where the six $a^\mu$ are the six J2000 Keplerian elements (semimajor axis, eccentricity, inclination, mean longitude, longitude of the perihelion, longitude of the ascending node), and the $\delta a^\mu$ are their perturbations. Alternatively, we can formulate the perturbations in terms of Brouwer and Clemence's ``Set III'' parameters \citep{1961mcm..book.....B}, for which we have access to uncertainty estimates for certain JPL SSEs (\citealt{de434,de438}; see Fig.\ \ref{fig:setIIIposteriors}). The two formulations are largely equivalent with respect to their effect on GW searches.

To implement the R{\o}mer-delay perturbations, we begin with approximate values for the $a^\mu$ and their rates of change.\footnote{The six $a^\mu$ plus the rate of change $\dot{a}^4 \equiv \dot{l}$ of mean longitude specify Jupiter's orbit as the osculating ellipse at the J2000 reference epoch (MJD 2451545); the remaining five rates encode the secular evolution of Jupiter's orbit due to SSB bodies other than the Sun, and to other physical effects. See \href{https://ssd.jpl.nasa.gov/txt/aprx_pos_planets.pdf}{https://ssd.jpl.nasa.gov/txt/aprx\_pos\_planets.pdf}.}
We vary these $a^\mu$ to minimize the (root-mean-square) difference between our quasi-Keplerian orbits and DE436, integrated between years 2000 and 2020. The resulting orbits are within 1\% of DE436, which ensures similar accuracy for the orbit partials, more than enough for our purposes.
\begin{figure*}[t]
    \centering
    \includegraphics[width=2\columnwidth]{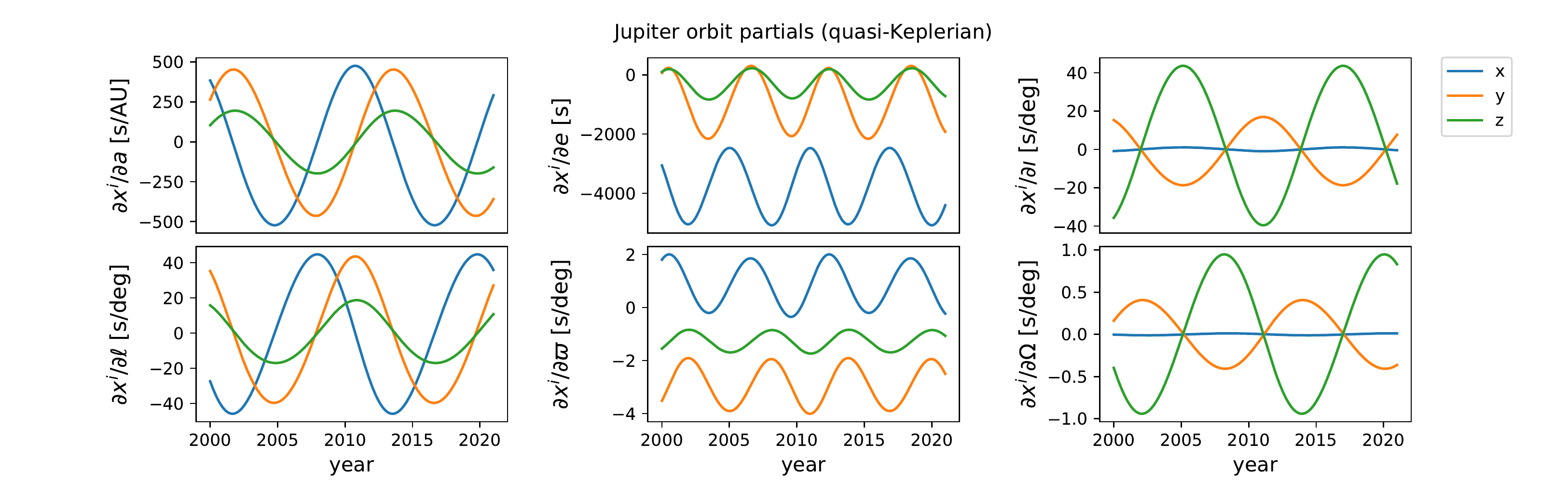}
    \includegraphics[width=2\columnwidth]{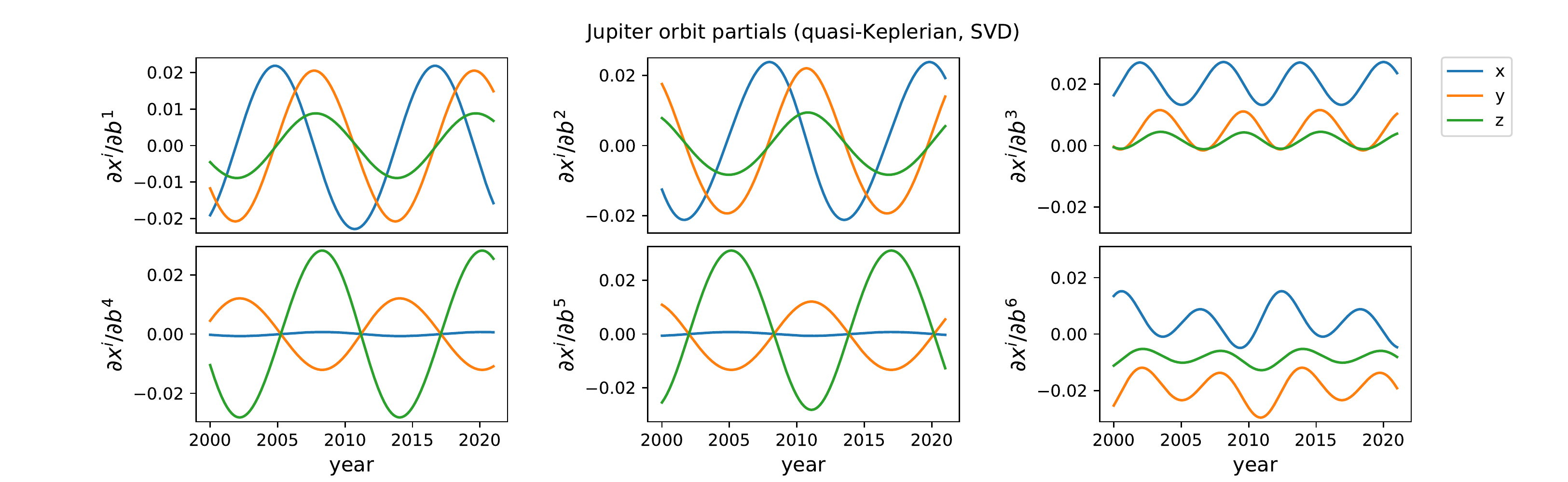}
    \caption{\textbf{Top}: perturbative partials ($x$, $y$, and $z$ coordinates, geometrized units of seconds) for Jupiter's orbit between the years 2000 and 2020, as obtained by varying the six Keplerian elements in a quasi-Keplerian model that includes also their rates. The baseline elements and rates are adjusted to fit DE436 orbits.
    Here $a$ is the semimajor axis of the orbit, $e$ the eccentricity, $\imath$ the inclination, $\ell$ the mean longitude, $\varomega$ the longitude of the periapsis, and $\Omega$ the longitude of the ascending node. 
    \textbf{Bottom}: singular-value-decomposition vector time series obtained from the partials in the top panel.}
    \label{fig:jupiterorbit}
\end{figure*}

We compute the six partials as finite differences; since they are strongly correlated (see top panel of Fig.\ \ref{fig:jupiterorbit}) and have different scales in their natural units, we decorrelate and normalize them by computing the singular value decomposition $\sum_\nu U_{\mu\nu} S_\nu V_{\nu i k}$ of the matrix $P_{\mu i k} = \partial x^{(5)}_k(t_i)/\partial a^\mu$, and adopting $\partial x^{(5)}_k(t_i)/\partial b^\nu \equiv V_{\nu i k}$ as new orbit partials (see bottom panel of Fig.\ \ref{fig:jupiterorbit}), where $\delta b^\nu = \sum_\mu \delta a^\mu M_{\mu \nu}$ with $M_{\mu \nu} = U_{\mu\nu} S_\nu$ (no summation intended). The resulting units are mixed. We give the orthonormalized coefficients $\delta b^\mu$ uniform priors that are broad enough to generate the range of R{\o}mer variation seen across the JPL SSEs, and to contain the support of the PTA likelihood for the NANOGrav 11-yr dataset. In other words, we make the priors broad enough that the posteriors do not impinge on the boundaries.

We do not include Uranus and Neptune's orbital perturbations, which lead to $\delta t_{\mathrm{R\odot}}$ with linear time dependencies that are absorbed entirely by timing-model parameters. By contrast, we repeat the procedure that we just outlined for Saturn, but find that GW posteriors are barely affected for orbital perturbations of magnitude comparable to the differences among the JPL SSEs. Thus in our use of \textsc{BayesEphem} we usually omit Saturn orbit perturbations. These may prove more important as datasets increase in length.
\end{itemize}

The components outlined above are brought together into a linear delay model $\delta t_{\mathrm{R}\odot}(c^a)$ written as the product of the eleven-dimensional correction vector $c^a \equiv \{\delta m^{(p)}, \omega^{\hat{z}}, \delta b^\nu\}$ (with $(p) = 5,6,7,8$ and $\nu = 1, \ldots, 6$) by the $(n_\mathrm{TOAs} \times 11)$-dimensional design matrix $G_{i a}$ with columns defined by the equations in this Section. This model, including variational partials for Keplerian and set-III parametrizations, is available as part of the open-source software package \textsc{Enterprise} \citep{2019ascl.soft12015E}.

It is also possible to treat the linear model as a Gaussian process common to all pulsars, and to marginalize analytically over the $c^a$, by building the effective correlation matrix $G_{i a} \Phi_{ab} G_{j b}$, where $\Phi_{ab}$ is prior covariance of the $c^a$ \citep{vhv14}. The basis vector for each $c^a$ corresponds to the concatenation of the R{\o}mer-delay perturbations that the parameter generates for each pulsar (i.e., the projection of the same vector time series onto different $\mathbf{p}$, at the appropriate TOAs).

\newpage 
\section{Effect of \textsc{BayesEphem} on the search for stochastic gravitational waves in NANOGrav's 11-yr dataset}
\label{sec:results}

To derive the SSE-robust results reported in \cite{2018ApJ...859...47A}, we sampled the \textsc{BayesEphem} corrections $c^a$ alongside the hyperparameters that describe the common GW background and the noise parameters for each pulsar (see Sec.\ 3.4 of \citealt{2018ApJ...859...47A} and Secs.\ 3 and 4 of \citealt{2016ApJ...821...13A}), treating $\delta t_{\mathrm{R}\odot}(c^a)$ as a deterministic correction to the residuals.

In Fig.\ \ref{fig:posteriors} we show Bayesian posteriors for the stochastic-background amplitude $A_\mathrm{GWB}$ at a fiducial frequency of yr$^{-1}$, as computed by fixing the fiducial SSE to DE421, DE430, DE435, DE436, and DE438 in turn, and either disabling (dotted lines) or applying \textsc{BayesEphem} in the Set III formulation (solid lines). The prior on $\log_{10} A_\mathrm{GWB}$ is flat in $[-18,-14]$, and the GWB spectral slope is set to $\gamma = 13/3$, as appropriate for an ensemble of binary inspirals progressing by GW emission alone. The posteriors follow from PTA likelihoods that omit Hellings--Downs correlations (as in ``model 2A'' rather than ``3A'' of \citealt{2018ApJ...859...47A}), but results change only modestly if we include those.
(Simulations show that as PTA datasets become more sensitive to GWs, a GWB would manifest first as seemingly uncorrelated red-noise processes of comparable amplitude in multiple pulsars, and later as a Hellings--Downs-correlated process across the PTA, providing more conclusive evidence of the GW origin of the signal.)

The plots in Fig.\ \ref{fig:posteriors} demonstrate the successful bridging of $A_\mathrm{GWB}$, our goal in mitigating the systematic effects of SSE errors. In doing so, \textsc{BayesEphem} removes hints that any GWB is present, as confirmed by computing model-2A Bayes factors (in favor of a common GWB process), listed in Table \ref{tab:bf}. In Table \ref{tab:upper} we show 95\% $A_\mathrm{GWB}$ upper limits, computed as just discussed but with flat uninformative priors on $A_\mathrm{GWB}$. The tables (as well as plots analogous to Fig.\ \ref{fig:posteriors} not shown here) confirm that the Set III and Keplerian-element formulations are equivalent with respect to GW detection.
\begin{table}[ht]
    \begin{center}
    \caption{Bayesian evidence for a GWB with different SSEs.\label{tab:bf}} 
    \begin{tabular}{l|r|r|r}
SSE & no \textsc{BayesEphem} & Set III & Keplerian \\
\hline
DE421 & 10.6 & 0.68 & 0.68 \\
DE430 & 23.7 & 0.72 & 0.71 \\
DE435 &  2.0 & 0.76 & 0.72 \\
DE436 &  6.2 & 0.80 & 0.72 \\
DE438 & 40.7 & 0.91 & 0.87
    \end{tabular}
    \tablecomments{Savage--Dickey Bayes factors in favor of a common $\gamma=13/3$ red-noise process (``model 2A'' of a GWB in \cite{2018ApJ...859...47A}) in NANOGrav's 11-yr dataset, as obtained by fixing the fiducial SSE shown in column 1, and omitting (column 2) or applying \textsc{BayesEphem} in its Set III (column 3) or Keplerian-element (column 4) formulations. The numbers shown here were drawn from newly reproduced Monte Carlo runs, and differ from those of \cite{2018ApJ...859...47A} by small sampling errors.}
\end{center}
\end{table}
\begin{table}[ht]
    \begin{center}
    \caption{95\% Bayesian upper limits for $A_\mathrm{GWB}$.\label{tab:upper}}
    \begin{tabular}{l|c|c|c}
SSE & no \textsc{BayesEphem} & Set III & Keplerian \\
\hline
DE421 & $1.54 \times 10^{-15}$ & $1.32 \times 10^{-15}$ & $1.35 \times 10^{-15}$ \\
DE430 & $1.76 \times 10^{-15}$ & $1.31 \times 10^{-15}$ & $1.33 \times 10^{-15}$ \\
DE435 & $1.59 \times 10^{-15}$ & $1.38 \times 10^{-15}$ & $1.40 \times 10^{-15}$ \\
DE436 & $1.64 \times 10^{-15}$ & $1.38 \times 10^{-15}$ & $1.41 \times 10^{-15}$ \\
DE438 & $1.94 \times 10^{-15}$ & $1.45 \times 10^{-15}$ & $1.44 \times 10^{-15}$ \\
    \end{tabular}
    \tablecomments{95\% Bayesian upper limits on the amplitude of a $\gamma=13/3$ GWB in NANOGrav's 11-yr dataset, listed as in Table \ref{tab:bf}. The numbers shown here were drawn from newly reproduced Monte Carlo runs, and differ from those of \cite{2018ApJ...859...47A} by sampling errors $\sim 0.02$.}
    \end{center}
\end{table}
\begin{figure*}[t]
    \includegraphics[width=2\columnwidth]{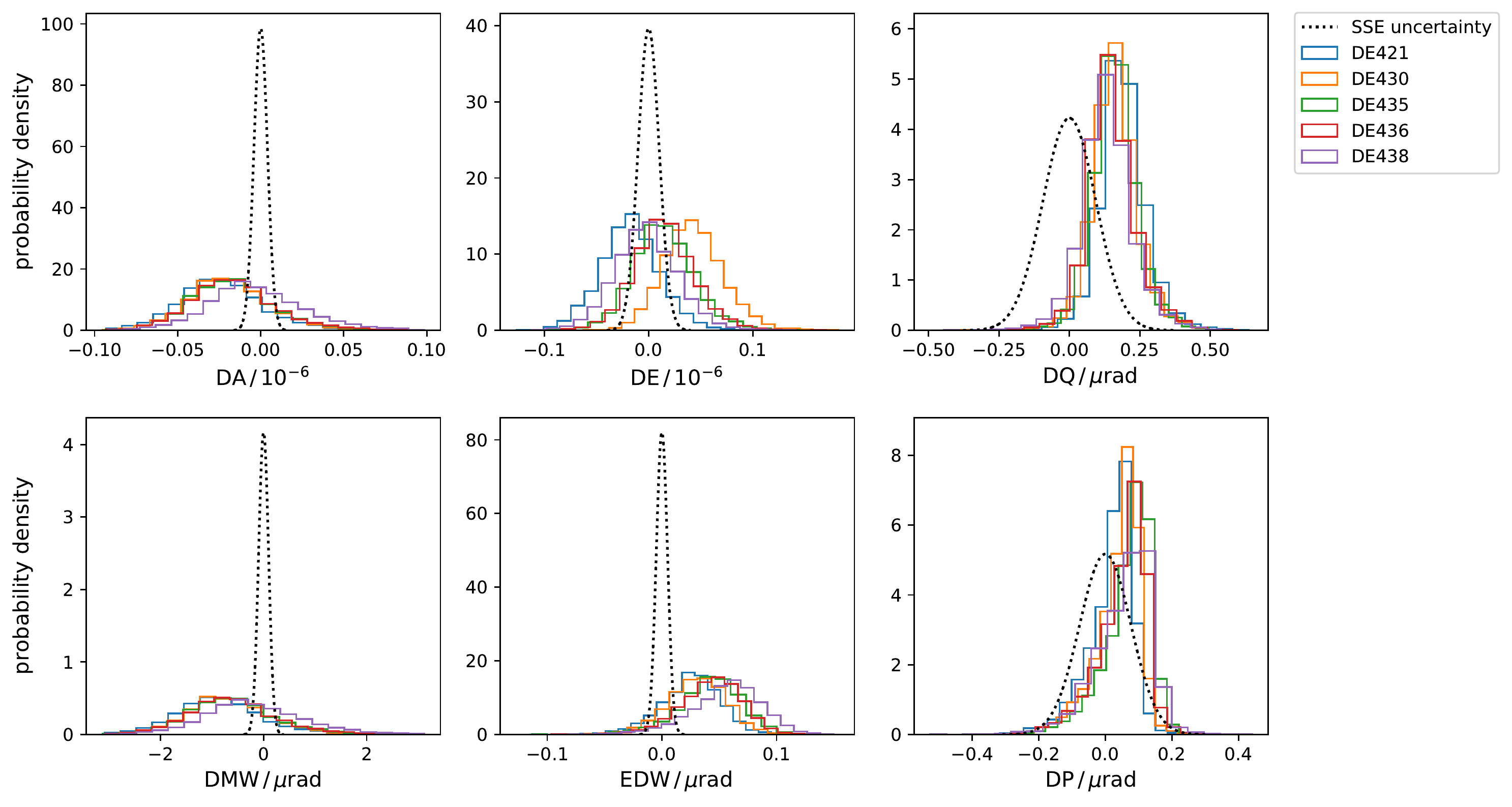}
    \caption{Bayesian posteriors for corrections to Jupiter's ``Set III'' orbital elements, as obtained in \textsc{BayesEphem}'s application to NANOGrav's 11-yr dataset. The solid curves show posteriors obtained by correcting different fiducial SSEs; the dotted curve shows JPL's estimate of uncertainties in Jupiter's osculating orbital elements for DE435 and DE436 \citep{de434}.
    Parameters names follow JPL's orbit-determination package \citep{moyer2003}. 
    Here $\mathrm{DA} \equiv \Delta a/a$, a fractional correction to the semimajor axis of the osculating orbit; $\mathrm{DE} \equiv \Delta e$, a fractional correction to the eccentricity; $\mathrm{DMW} = \Delta M_0 + \Delta w$, $\mathrm{EDW} = e \Delta w$, $\mathrm{DP} = \Delta p$, and $\mathrm{DQ} = \Delta q$, where $M_0$ is the mean anomaly at the initial epoch (conventionally MJD 2440400.5) and $(\Delta p, \Delta q, \Delta w)$ encode a rigid rotation of the orbit, with $\Delta p$ and $\Delta q$ in the orbital plane, and $\Delta w$ normal to it.}
    \label{fig:setIIIposteriors}
\end{figure*}

The posterior distributions of the \textsc{BayesEphem} mass perturbations are identical to their IAU priors:  NANOGrav's 11-yr dataset is uninformative compared to current estimates from spacecraft data. Likewise, the frame rotation rate $\omega^{\hat{z}}$ fills its prior, as expected.
The posterior distributions of the \textsc{BayesEphem} corrections to Jupiter's orbital elements are shown in Fig.\ \ref{fig:setIIIposteriors}, where they are compared to JPL's estimated uncertainties for DE435 and DE436, derived by comparing SSE fits that use independent subsets of the data \citep{de434}. The NANOGrav posteriors appear consistent with all SSEs: they have reasonably high support around the estimated-uncertainty region around $\delta a^\mu = 0$. The least consistent parameters are those involving rotations of the orbital plane (DMW and EDW, see the Fig.\ \ref{fig:setIIIposteriors} caption), as well as the eccentricity (DE) for the oldest SSEs, DE421 and DE430. JPL's estimated uncertainties for DE438 are a factor of $\sim$ four tighter than for DE435/6, but they do not change this picture substantially.

The large dispersion of the \textsc{BayesEphem} posteriors is not unexpected, given that pulsar-timing data is sensitive to R{\o}mer delays in a rather selective fashion (see the discussion of Sec.\ \ref{sec:systematics}).
In terms of uncertainties on Jupiter's instantaneous position, the \textsc{BayesEphem} posteriors map to RMS 3D errors $\sim 100$ km, thus larger than the differences between SSEs. While \textsc{BayesEphem} can bridge NANOGrav 11-yr $A_\mathrm{GW}$ posteriors for different SSEs, the resulting orbital-element posteriors 
do not identify any specific systematic offset among them.

\section{Assessing gravitational-wave-background detection prospects when using \textsc{BayesEphem}}
\label{sec:simulations}

\textsc{BayesEphem} is designed to model SSE uncertainties \emph{and} systematics, bridging published estimates of Earth's trajectory around the SSB.
To do so, \textsc{BayesEphem} introduces new parameters that govern a spatially correlated process of amplitude comparable to the stochastic GWBs that we seek.
While SSE corrections and the GWB have different spatial-correlation structures, the two may nevertheless remain degenerate when probed by a limited number of PTA pulsars \citep{2019ApJ...876...55R}.
It is then natural to ask how the application of \textsc{BayesEphem} may affect GWB detection prospects (sensitivity and time to detection) in the weak-to-intermediate signal-to-noise regime in which spatial correlations among the pulsars carry marginal information.
To sketch an answer to this question we conducted a set of simulations, which are summarized briefly in \cite{2018ApJ...859...47A} and discussed further here.

To wit, we created multiple synthetic datasets meant to replicate the sensitivity of NANOGrav's 11-yr (really, 11.4-yr) data. We used actual observation epochs for the 34 analyzed pulsars, and extended the time span to 15 years by drawing observation times from distributions fit to the last three years of measured data.
We simulated timing residuals by drawing random white- and red-noise deviates at the maximum-a-posteriori 11-yr levels, and again extrapolated to 15 years from the last three of the 11.
These choices ensure that future data taking is represented with cadence and precision comparable to the end of 2015---a very conservative option given ongoing receiver upgrades and the continuous addition of new pulsars to the PTA.

We calibrated these noise-only simulations by increasing noise levels until the 11-yr ``slice'' of the data matched the GWB upper limit (specifically the spatially uncorrelated model 2A, with uncorrected DE436) of \cite{2018ApJ...859...47A}.
All simulated datasets were created by postulating that DE436 was the ``truth.'' We then injected $\gamma = 13/3$ GWBs of various amplitudes, and analyzed both the 11-yr slices and the full 15-yr datasets by taking DE430 as our fiducial SSE, both with and without \textsc{BayesEphem}.
Our results were as follows.

First, for noise-only datasets without \textsc{BayesEphem}, the systematic offset between DE430 and DE436 is interpreted as a GWB for both 11-yr and 15-yr time spans, with (model 2A vs.\ the noise-only model 1) Bayes ratios of $\sim 2$ and $\sim 20$ respectively. The ratios are reduced to levels consistent with noise fluctuations by the application of \textsc{BayesEphem}.
\begin{figure}
    \includegraphics{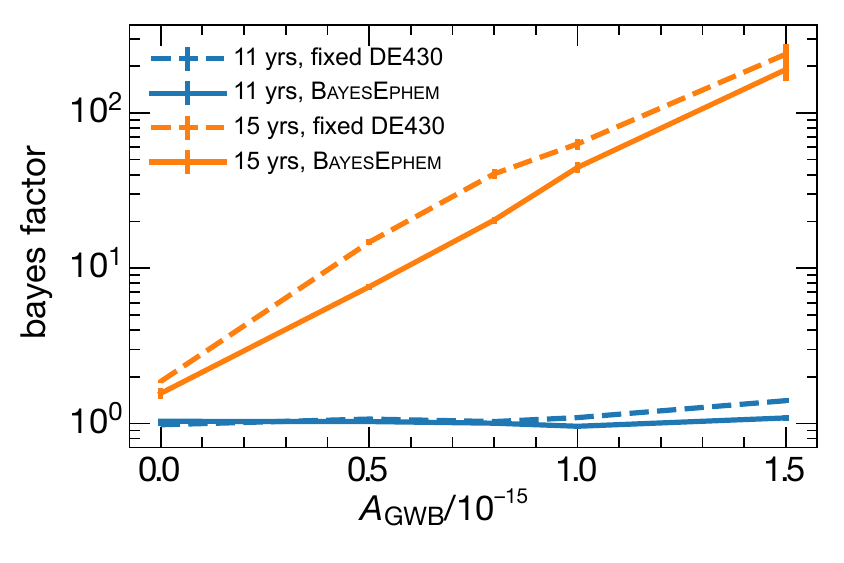}
    \caption{Correlated- vs.\ uncorrelated-GWB Bayes factors (i.e., model 3A vs.\ model 2A), estimated for a set of simulated $15$-yr datasets with GWB injections at increasing amplitudes.
    The datasets were generated by adopting DE436 as the ``correct'' SSE, and analyzed using both the ``wrong'' DE430 ephemeris (dashed lines) and \textsc{BayesEphem} (solid). The orange and blue lines display results for the full datasets and their 11-yr ``slices,'' respectively.
    Their comparison indicates that \textsc{BayesEphem} will not impede the ability of PTAs to make a definitive detection in the near future.
    Figure adapted from \cite{2018ApJ...859...47A}.
    \label{fig:simulations}}
\end{figure}

Second, as we increase the amplitude of injected GWBs while applying \textsc{BayesEphem}, the Bayes factors remain marginal for the 11-yr dataset, as do model 3A-vs.-model 2A factors, the substantive spatial-correlation test of GW presence. The latter are shown in Fig.\ \ref{fig:simulations}.
The story is different for 15-yr datasets, where the scaling of Bayes factors with injected GWBs is comparable whether or not we apply \textsc{BayesEphem}. Indeed, the longer time span disentangles SSB and GWB correlations, enabling detection even at the astrophysically conservative levels described by Sesana and colleagues \citep{2016MNRAS.463L...6S}.

In summary, our simulations confirm that SSE errors can produce spurious evidence for GWBs; they suggest that, in datasets similar to NANOGrav's 11-yr, \textsc{BayesEphem} can overcorrect these biases, suppressing the evidence for a ``true'' GWB; but they imply also that this effect vanishes for longer datasets. Thus, we expect that \textsc{BayesEphem} will not impair detection prospects in the near future---if of course nature grants us a sufficient GWB bounty.

\section{Other modeling approaches}
\label{sec:othermodels}

In our analysis of NANOGrav's 11-yr dataset, we experimented with other modeling approaches for SSE uncertainties, which (in our implementation) did not satisfy our bridging criterion, or were otherwise disfavored. We discuss them briefly here, as a reference for future investigations:
\begin{itemize} 
    \item \textbf{Planetary mass perturbations for outer-SS planets:} modeled as in \cite{2010ApJ...720L.201C} (and as in the first element of \textsc{BayesEphem}, see Sec.\ \ref{sec:physical}), these did not affect our $A_\mathrm{GWB}$ posteriors significantly (and therefore did not resolve the discrepancy among SSEs), whether introduced with best-estimate priors \citep{iaumasses} or with more relaxed assumptions.
    \item \textbf{Dipole-correlated Gaussian process:}
    as proposed in \cite{2016MNRAS.455.4339T}, the R{\o}mer delays for individual pulsars may be treated as Gaussian processes with common power-law or free-spectral priors \citep{vhv14}, and with dipolar spatial correlations between pulsars ($C_{ij} = \cos \theta_{ij}$, with $\theta_{ij}$ the angle between pulsars).
    We note that this approach is equivalent to modeling the apparent motion of the SSB (or equivalently, the error in Earth's orbit) as three Gaussian processes (along the three spatial axes) with identical uncorrelated priors, and then projecting the vector time series onto the pulsar positions to obtain R{\o}mer delays.
    
    Dipole-correlated Gaussian processes were included in models 2B, 2C, 3B, and 3C of our 11-yr analysis \citep{2018ApJ...859...47A}.
    It is unclear to us why the approach failed to bridge $A_\mathrm{GWB}$ posteriors across SSEs, but the reason may be related to the choice of Gaussian-process priors appropriate to describe the range and shapes of variations among SSEs.
    \item \textbf{R{\o}mer mixture:} this phenomenological model describes Earth's orbit as a linear combination of its estimate in multiple SSEs, with mixture coefficients constrained by a Dirichlet prior (see, e.g., \citealt{gelman2013bayesian}). 
    This approach achieves bridging by construction, but it is difficult to interpret physically.
    In our 11-yr investigation, the posteriors of the mixture coefficients indicated a moderate preference for DE435/6 over DE421 and DE430.
    \item \textbf{Gaussian process based on numerical partials:} in this approach, SSE corrections are modeled as a finite Gaussian process \citep{williams2006gaussian} in which basis vectors are given by orbit partials (e.g., the change in Earth's orbit as we vary the initial conditions for all the planets), as computed by numerical integration for SSE fits. The priors for the basis weights is given by the formal covariance of the SSE fit parameters. The basis vectors are then projected into R{\o}mer delays for each pulsar.
    
    It turns out that only Jupiter and Saturn partials matter to GW results; when restricted to these planets, the approach is effectively equivalent to the orbit-correction sector of \textsc{BayesEphem}---at least for the 11-yr time span, for which numerical partials are very close to analytical osculating-orbit perturbations.
\end{itemize}

\section{Conclusion}
\label{sec:conclusion}

It is striking that the abundance and precision of current PTA datasets should be such that our sensitivity to GWs is limited by the very accuracy to which we can position Earth around the SSB.
In the analysis of NANOGrav's 11-yr dataset \citep{2018ApJ...859...47A}, we took the conservative position that robust GW upper limits and detection Bayes factors must be reproducible with all SSEs released in the last ten years, after introducing a sufficiently descriptive model of SSE uncertainties.
To this end, we developed \textsc{BayesEphem}, described in this article, which focuses on the SSE degrees of freedom (Jupiter's orbital elements) that measurably affect our GW search, and which produces ``bridged'' $A_\mathrm{GWB}$ posteriors, upper limits, and Bayes factors. Because of this focus, the analysis of NANOGrav data does not update the JPL SSEs significantly (see Fig.\ \ref{fig:setIIIposteriors}), although it provides uncertainty estimates entirely independent of those offered by SSE makers \citep{de434,de438}. We expect that it would be possible to concentrate instead on a sector of SSE corrections that do not affect GW results, but that would benefit from PTA constraints, thus producing PTA-enhanced SSEs useful beyond GW detection. For this purpose we could adopt either the orbital-elements formalism of Sec.\ \ref{sec:physical} or the numerical-partials approach of Sec.\ \ref{sec:othermodels}.

The conservative modeling attitude employed for NANOGrav's 11-yr analysis comes at the cost of a loss of GW sensitivity, as described in Sec.\ \ref{sec:simulations} (see also \citealt{2019ApJ...876...55R}). The GW statistics reported in \cite{2018ApJ...859...47A} (and here in Tables \ref{tab:bf} and \ref{tab:upper}) are supported in our Bayesian setting, in which we decline to favor one SSE above others; but these results should not be considered binding for future GW searches that rely on demonstrably accurate SSEs and that are based on longer datasets where SSE and GW correlations become disentangled:
more precisely, datasets that cover a longer timespan with a sufficient number of high-quality pulsars.
Both conditions are now materializing: Juno's ongoing measurements are improving estimates of Jupiter's orbit \citep{de438}, which (we argue) is the limiting factor for GW searches; and the NANOGrav dataset is progressing toward the 15-yr span, for which (we reckon) jovian systematics decouple from GW statistics.
The combined datasets assembled by the International Pulsar Timing Array \citep{2019MNRAS.490.4666P} have already passed this mark and thus may already be immune to this problem.

The path toward an authoritative detection of low-frequency GWs with PTAs requires intense and persistent timing campaigns on the world's most sensitive radiotelescopes; 
sophisticated inference techniques and clever high-performance-computing algorithms, to make sense of ever-growing, heterogeneous datasets with many unknown parameters; and a confident control of TOA systematics at the ns level. Among the last, the error analysis and validation of high-precision SSEs will remain paramount.

\begin{center}
\rule{0.25\columnwidth}{.4pt}
\end{center}

\emph{Author contributions.} 
This document is the result of more than a decade of work by the entire NANOGrav collaboration.
We acknowledge specific contributions below.
ZA, KC, PBD, MED, TD, JAE, RDF, ECF, EF, PAG, GJ, MLJ, MTL, LL, DRL, RSL, MAM, CN, DJN, TTP, SMR, PSR, RS, IHS, KS, JKS, and WWZ developed the 11-year data set.
MV\ led this analysis and coordinated writing.
MV, SRT, and JS developed \textsc{BayesEphem}, managed Bayesian-inference runs, and performed data analysis, with help from CC, JAE, TJWL, and SJV.
MV, SRT, and JS wrote this article, with edits by DJN, MTL, TJWL, DLK, JSH, and SJV.
Ephemeris data and expertise were provided by WMF and RSP.

\\

\emph{Acknowledgments.} 
We are grateful to Agn\`es Fienga, Joseph Romano, Alvin Chua, and Maria Charisi for useful comments and interactions.

The NANOGrav project receives support from National Science Foundation (NSF) Physics Frontier Center award \#1430284.
NANOGrav research at UBC is supported by an NSERC Discovery Grant and Discovery Accelerator Supplement and by the Canadian Institute for Advanced Research.
M.V.\ and J.S.\ acknowledge support from the JPL RTD program.
S.R.T.\ was partially supported by an appointment to the NASA Postdoctoral Program at JPL, administered by Oak Ridge Associated Universities through a contract with NASA.
J.A.E.\ was partially supported by NASA through Einstein Fellowship grants PF4-150120.
Portions of this work performed at NRL are supported by the Chief of Naval Research.
The Flatiron Institute is supported by the Simons Foundation.
Portions of this research were carried out at the Jet Propulsion Laboratory, California Institute of Technology, under a contract with the National Aeronautics and Space Administration.

This work was supported in part by National Science Foundation Grant PHYS-1066293 and by the hospitality of the Aspen Center for Physics.
We are grateful for computational resources provided by the Leonard E.\ Parker Center for Gravitation, Cosmology and Astrophysics at the University of Wisconsin-Milwaukee, which is supported by NSF Grants 0923409 and 1626190.
Data for this project were collected using the facilities of the Green Bank Observatory and the Arecibo Observatory.
The National Radio Astronomy Observatory and Green Bank Observatory is a facility of the National Science Foundation operated under cooperative agreement by Associated Universities, Inc.
The Arecibo Observatory is a facility of the National Science Foundation operated under cooperative agreement by the University of Central Florida in alliance with Yang Enterprises, Inc.\ and Universidad Metropolitana.
Copyright 2020. All rights reserved.

\begin{center}
\rule{0.25\columnwidth}{.4pt}
\end{center}

\bibliographystyle{yahapj}
\bibliography{references}

\begin{thebibliography}{}
\providecommand\natexlab[1]{#1}
\providecommand\JournalTitle[1]{#1}

\bibitem[{{Arzoumanian} {et~al.}(2016)}]{2016ApJ...821...13A}
{Arzoumanian}, Z., {et~al.} 2016,
  \href{http://dx.doi.org/10.3847/0004-637X/821/1/13}{\JournalTitle{\apj}, 821,
  13}

\bibitem[{{Arzoumanian} {et~al.}(2018{\natexlab{a}})}]{2018ApJS..235...37A}
---. 2018{\natexlab{a}},
  \href{http://dx.doi.org/10.3847/1538-4365/aab5b0}{\JournalTitle{Astrophys. J.
  Supp.}, 235, 37}

\bibitem[{{Arzoumanian} {et~al.}(2018{\natexlab{b}})}]{2018ApJ...859...47A}
---. 2018{\natexlab{b}},
  \href{http://dx.doi.org/10.3847/1538-4357/aabd3b}{\JournalTitle{\apj}, 859,
  47}

\bibitem[{{Brouwer} \& {Clemence}(1961)}]{1961mcm..book.....B}
{Brouwer}, D., \& {Clemence}, G.~M. 1961, {Methods of celestial mechanics} (New
  York: Academic Press)

\bibitem[{{Caballero} {et~al.}(2018)}]{2018MNRAS.481.5501C}
{Caballero}, R.~N., {et~al.} 2018,
  \href{http://dx.doi.org/10.1093/mnras/sty2632}{\JournalTitle{MNRAS}, 481,
  5501}

\bibitem[{{Champion} {et~al.}(2010)}]{2010ApJ...720L.201C}
{Champion}, D.~J., {et~al.} 2010,
  \href{http://dx.doi.org/10.1088/2041-8205/720/2/L201}{\JournalTitle{\apj},
  720, L201}

\bibitem[{{Desvignes} {et~al.}(2016)}]{dcl+16}
{Desvignes}, G., {et~al.} 2016,
  \href{http://dx.doi.org/10.1093/mnras/stw483}{\JournalTitle{MNRAS}, 458,
  3341}

\bibitem[{{Detweiler}(1979)}]{det79}
{Detweiler}, S. 1979,
  \href{http://dx.doi.org/10.1086/157593}{\JournalTitle{Astrophys. J.}, 234,
  1100}

\bibitem[{Dickey(1971)}]{d71}
Dickey, J.~M. 1971,
  \href{http://www.jstor.org/stable/2958475}{\JournalTitle{The Annals of
  Mathematical Statistics}, 42, 204}

\bibitem[{{Edwards} {et~al.}(2006{\natexlab{a}}){Edwards}, {Hobbs}, \&
  {Manchester}}]{2006MNRAS.372.1549E}
{Edwards}, R.~T., {Hobbs}, G.~B., \& {Manchester}, R.~N. 2006{\natexlab{a}},
  \href{http://dx.doi.org/10.1111/j.1365-2966.2006.10870.x}{\JournalTitle{MNRAS},
  372, 1549}

\bibitem[{{Edwards} {et~al.}(2006{\natexlab{b}}){Edwards}, {Hobbs}, \&
  {Manchester}}]{ehm06}
---. 2006{\natexlab{b}},
  \href{http://dx.doi.org/10.1111/j.1365-2966.2006.10870.x}{\JournalTitle{MNRAS},
  372, 1549}

\bibitem[{{Ellis} {et~al.}(2017){Ellis}, {Vallisneri}, {Taylor}, \&
  {Baker}}]{enterprise}
{Ellis}, J.~A., {Vallisneri}, M., {Taylor}, S.~R., \& {Baker}, P.~T. 2017,
  Enterprise, \url{https://github.com/nanograv/enterprise}

\bibitem[{{Fey} {et~al.}(2015)}]{2015AJ....150...58F}
{Fey}, A.~L., {et~al.} 2015,
  \href{http://dx.doi.org/10.1088/0004-6256/150/2/58}{\JournalTitle{Astron.
  J.}, 150, 58}

\bibitem[{Fienga {et~al.}(2014)Fienga, Manche, Laskar, Gastineau, \&
  Verma}]{inpop13}
Fienga, A., Manche, H., Laskar, J., Gastineau, M., \& Verma, A. 2014, INPOP new
  release: INPOP13b, \href{http://arxiv.org/abs/1405.0484}{{\sffamily
  arXiv:1405.0484 [astro-ph.EP]}}

\bibitem[{Folkner \& Park(2015)}]{de434cov}
Folkner, W.~M., \& Park, R.~S. 2015, Uncertainty in the orbit of Jupiter for
  Juno planning, Tech. Rep. IOM 343R-15-019, Jet Propulsion Laboratory,
  Pasadena, CA

\bibitem[{Folkner \& Park(2016)}]{de436}
---. 2016, {JPL planetary and Lunar ephemeris DE436}, online,
  \url{https://naif.jpl.nasa.gov/pub/naif/JUNO/kernels/spk/de436s.bsp.lbl}

\bibitem[{Folkner \& Park(2018)}]{de438}
---. 2018, Planetary ephemeris DE438 for Juno, Tech. Rep. IOM 392R-18-004, Jet
  Propulsion Laboratory, Pasadena, CA

\bibitem[{Folkner {et~al.}(2016)Folkner, Park, \& Jacobson}]{de435}
Folkner, W.~M., Park, R.~S., \& Jacobson, R.~A. 2016, Planetary ephemeris
  DE435, Tech. Rep. IOM 392R-16-003, Jet Propulsion Laboratory, Pasadena, CA,
  \url{ftp://ssd.jpl.nasa.gov/pub/eph/planets/ioms/de435.iom.pdf}

\bibitem[{{Folkner} {et~al.}(2009){Folkner}, {Williams}, \&
  {Boggs}}]{2009IPNPR.178C...1F}
{Folkner}, W.~M., {Williams}, J.~G., \& {Boggs}, D.~H. 2009,
  \href{https://ipnpr.jpl.nasa.gov/progress_report/42-178/178C.pdf}{\JournalTitle{Interplanetary
  Network Progress Report}, 178, 1}

\bibitem[{{Folkner} {et~al.}(2014){Folkner}, {Williams}, {Boggs}, {Park}, \&
  {Kuchynka}}]{2014IPNPR.196C...1F}
{Folkner}, W.~M., {Williams}, J.~G., {Boggs}, D.~H., {Park}, R.~S., \&
  {Kuchynka}, P. 2014,
  \href{https://naif.jpl.nasa.gov/pub/naif/generic_kernels/spk/planets/de430_and_de431.pdf}{\JournalTitle{Interplanetary
  Network Progress Report}, 196, 1}

\bibitem[{{Foster} \& {Backer}(1990{\natexlab{a}})}]{1990ApJ...361..300F}
{Foster}, R.~S., \& {Backer}, D.~C. 1990{\natexlab{a}},
  \href{http://dx.doi.org/10.1086/169195}{\JournalTitle{\apj}, 361, 300}

\bibitem[{{Foster} \& {Backer}(1990{\natexlab{b}})}]{fb90}
---. 1990{\natexlab{b}},
  \href{http://dx.doi.org/10.1086/169195}{\JournalTitle{Astrophys. J.}, 361,
  300}

\bibitem[{Gelman {et~al.}(2013)Gelman, Carlin, Stern, Dunson, Vehtari, \&
  Rubin}]{gelman2013bayesian}
Gelman, A., Carlin, J.~B., Stern, H.~S., {et~al.} 2013, Bayesian data analysis
  (Chapman and Hall/CRC)

\bibitem[{{Guo} {et~al.}(2019){Guo}, {Li}, {Lee}, \&
  {Caballero}}]{2019MNRAS.489.5573G}
{Guo}, Y.~J., {Li}, G.~Y., {Lee}, K.~J., \& {Caballero}, R.~N. 2019,
  \href{http://dx.doi.org/10.1093/mnras/stz2515}{\JournalTitle{MNRAS}, 489,
  5573}

\bibitem[{Hees {et~al.}(2014)Hees, Folkner, Jacobson, \&
  Park}]{PhysRevD.89.102002}
Hees, A., Folkner, W.~M., Jacobson, R.~A., \& Park, R.~S. 2014,
  \href{http://dx.doi.org/10.1103/PhysRevD.89.102002}{\JournalTitle{Phys. Rev.
  D}, 89, 102002}

\bibitem[{{Hellings} \& {Downs}(1983)}]{hd83}
{Hellings}, R.~W., \& {Downs}, G.~S. 1983,
  \href{http://dx.doi.org/10.1086/183954}{\JournalTitle{Astrophys. J. Lett.},
  265, L39}

\bibitem[{{Hobbs}(2013)}]{h13}
{Hobbs}, G. 2013,
  \href{http://dx.doi.org/10.1088/0264-9381/30/22/224007}{\JournalTitle{Classical
  and Quantum Gravity}, 30, 224007}

\bibitem[{{Hobbs} \& {Edwards}(2012)}]{2012ascl.soft10015H}
{Hobbs}, G., \& {Edwards}, R. 2012,
  \href{https://ascl.net/1210.015}{\JournalTitle{Astrophysics Source Code
  Library}, ascl:1210.015}

\bibitem[{IAU(2017)}]{iaumasses}
IAU. 2017, {Division I Working Group on Numerical Standards for Fundamental
  Astronomy}, online, \url{http://maia.usno.navy.mil/NSFA/NSFA\_cbe.html}

\bibitem[{{Jacobson}(2009)}]{2009AJ....137.4322J}
{Jacobson}, R.~A. 2009,
  \href{http://dx.doi.org/10.1088/0004-6256/137/5/4322}{\JournalTitle{Astron.
  J.}, 137, 4322}

\bibitem[{{Jacobson}(2014)}]{2014AJ....148...76J}
---. 2014,
  \href{http://dx.doi.org/10.1088/0004-6256/148/5/76}{\JournalTitle{Astron.
  J.}, 148, 76}

\bibitem[{Jacobson {et~al.}(2000)Jacobson, Haw, McElrath, \&
  Antreasian}]{jh+2000}
Jacobson, R.~A., Haw, R.~J., McElrath, T.~P., \& Antreasian, P.~G. 2000,
  \JournalTitle{J. Astronaut. Sci.}, 48, 495

\bibitem[{{Jacobson} {et~al.}(2006)}]{2006AJ....132.2520J}
{Jacobson}, R.~A., {et~al.} 2006,
  \href{http://dx.doi.org/10.1086/508812}{\JournalTitle{Astron. J.}, 132, 2520}

\bibitem[{{Lommen} \& {Demorest}(2013)}]{2013CQGra..30v4001L}
{Lommen}, A.~N., \& {Demorest}, P. 2013,
  \href{http://dx.doi.org/10.1088/0264-9381/30/22/224001}{\JournalTitle{Classical
  and Quantum Gravity}, 30, 224001}

\bibitem[{{Lorimer} \& {Kramer}(2012)}]{2012hpa..book.....L}
{Lorimer}, D.~R., \& {Kramer}, M. 2012, {Handbook of Pulsar Astronomy}
  (Cambridge, UK: Cambridge University Press)

\bibitem[{Ma {et~al.}(1998)Ma, Arias, Eubanks, Fey, Gontier, Jacobs, Sovers,
  Archinal, \& Charlot}]{Ma_1998}
Ma, C., Arias, E.~F., Eubanks, T.~M., {et~al.} 1998,
  \href{http://dx.doi.org/10.1086/300408}{\JournalTitle{The Astronomical
  Journal}, 116, 516}

\bibitem[{{McLaughlin}(2013)}]{ml13}
{McLaughlin}, M.~A. 2013,
  \href{http://dx.doi.org/10.1088/0264-9381/30/22/224008}{\JournalTitle{Classical
  and Quantum Gravity}, 30, 224008}

\bibitem[{Moyer(2003)}]{moyer2003}
Moyer, T.~D. 2003, Formulation for Observed and Computed Values of Deep Space
  Network Data Types for Navigation (New York: Wiley)

\bibitem[{{Newhall} {et~al.}(1983){Newhall}, {Standish}, \&
  {Williams}}]{1983A&A...125..150N}
{Newhall}, X.~X., {Standish}, E.~M., \& {Williams}, J.~G. 1983,
  \JournalTitle{A\&A}, 125, 150

\bibitem[{Park {et~al.}(2015)Park, Folkner, \& Jacobson}]{de434}
Park, R.~S., Folkner, W.~M., \& Jacobson, R.~A. 2015, The planetary ephemeris
  DE434, Tech. Rep. IOM 392R-15-018, Jet Propulsion Laboratory, Pasadena, CA,
  \url{ftp://ssd.jpl.nasa.gov/pub/eph/planets/ioms/de434.iom.pdf}

\bibitem[{{Perera} {et~al.}(2019)}]{2019MNRAS.490.4666P}
{Perera}, B.~B.~P., {et~al.} 2019,
  \href{http://dx.doi.org/10.1093/mnras/stz2857}{\JournalTitle{MNRAS}, 490,
  4666}

\bibitem[{{Roebber}(2019)}]{2019ApJ...876...55R}
{Roebber}, E. 2019,
  \href{http://dx.doi.org/10.3847/1538-4357/ab100e}{\JournalTitle{\apj}, 876,
  55}

\bibitem[{R\o{}mer(1676)}]{Roemer1676}
R\o{}mer, O.~C. 1676,
  \href{https://gallica.bnf.fr/ark:/12148/bpt6k56527v/f234.item}{\JournalTitle{Journal
  des S\c{c}avans}, 11, 233}

\bibitem[{{Sazhin}(1978)}]{saz78}
{Sazhin}, M.~V. 1978, \JournalTitle{Sov.~Ast.}, 22, 36

\bibitem[{{Sesana} {et~al.}(2016){Sesana}, {Shankar}, {Bernardi}, \&
  {Sheth}}]{2016MNRAS.463L...6S}
{Sesana}, A., {Shankar}, F., {Bernardi}, M., \& {Sheth}, R.~K. 2016,
  \href{http://dx.doi.org/10.1093/mnrasl/slw139}{\JournalTitle{MNRAS}, 463, L6}

\bibitem[{{Shannon} {et~al.}(2015)}]{2015Sci...349.1522S}
{Shannon}, R.~M., {et~al.} 2015,
  \href{http://dx.doi.org/10.1126/science.aab1910}{\JournalTitle{Science}, 349,
  1522}

\bibitem[{{Tiburzi} {et~al.}(2016){Tiburzi}, {Hobbs}, {Kerr}, {Coles}, {Dai},
  {Manchester}, {Possenti}, {Shannon}, \& {You}}]{2016MNRAS.455.4339T}
{Tiburzi}, C., {Hobbs}, G., {Kerr}, M., {et~al.} 2016,
  \href{http://dx.doi.org/10.1093/mnras/stv2143}{\JournalTitle{MNRAS}, 455,
  4339}

\bibitem[{{van Haasteren} \& {Vallisneri}(2014)}]{vhv14}
{van Haasteren}, R., \& {Vallisneri}, M. 2014,
  \href{http://dx.doi.org/10.1103/PhysRevD.90.104012}{\JournalTitle{\prd}, 90,
  104012}

\bibitem[{{Verbiest} {et~al.}(2016)}]{v+16}
{Verbiest}, J.~P.~W., {et~al.} 2016,
  \href{http://dx.doi.org/10.1093/mnras/stw347}{\JournalTitle{MNRAS}, 458,
  1267}

\bibitem[{Verma(2013)}]{verma2013}
Verma, A.~K. 2013, PhD thesis, Universit\'e de Franche-Comt\'e, Besan\c{c}on,
  France

\bibitem[{Viswanathan {et~al.}(2018)Viswanathan, Fienga, Minazzoli, Bernus,
  Laskar, \& Gastineau}]{inpop17}
Viswanathan, V., Fienga, A., Minazzoli, O., {et~al.} 2018,
  \href{http://dx.doi.org/10.1093/mnras/sty096}{\JournalTitle{Monthly Notices
  of the Royal Astronomical Society}, 476, 1877}

\bibitem[{Williams \& Rasmussen(2006)}]{williams2006gaussian}
Williams, C.~K., \& Rasmussen, C.~E. 2006, Gaussian processes for machine
  learning (Cambridge, MA: MIT press)

\end{thebibliography}

\end{document}